# Adaptive Meta-Learning Stochastic Gradient Hamiltonian Monte Carlo Simulation for Bayesian Updating of Structural Dynamic Models

Xianghao Meng[1,2], James L. Beck[3], Yong Huang[1,2,*], Hui Li[1,2]

[1] Key Lab of Smart Prevention and Mitigation of Civil Engineering Disasters of the Ministry of Industry and Information Technology, Harbin Institute of Technology, Harbin, China

[2] Key Lab of Structures Dynamic Behavior and Control of the Ministry of Education, Harbin Institute of Technology, Harbin, China

[3] Division of Engineering and Applied Science, California Institute of Technology, CA, USA

**Abstract**

In the last few decades, Markov chain Monte Carlo (MCMC) methods have been widely applied to Bayesian updating of structural dynamic models in the field of structural health monitoring. Recently, several MCMC algorithms have been developed that incorporate neural networks to enhance their performance for specific Bayesian model updating problems. However, a common challenge with these approaches lies in the fact that the embedded neural networks often necessitate retraining when faced with new tasks, a process that is time-consuming and significantly undermines the competitiveness of these methods. This paper introduces a newly developed adaptive meta-learning stochastic gradient Hamiltonian Monte Carlo (AM-SGHMC) algorithm. The idea behind AM-SGHMC is to optimize the sampling strategy by training adaptive neural networks, and due to the adaptive design of the network inputs and outputs, the trained sampler can be directly applied to various Bayesian updating problems of the same type of structure without further training, thereby achieving meta-learning. Additionally, practical issues for the feasibility of the AM-SGHMC algorithm for structural dynamic model updating are addressed, and two examples involving Bayesian updating of multi-story building models with different model fidelity are used to demonstrate the effectiveness and generalization ability of the proposed method.

## 1. INTRODUCTION

Bayesian inference [1,2] is a general, rational, and robust tool which is widely used for structural model updating in structural health monitoring (SHM) [3-7]. The model updating problem is treated using Bayes' theorem to determine the posterior distribution of the parameter vector based on the available data. However, multidimensional integrals are usually encountered when calculating the posterior distribution, making analytical calculation intractable [1,2]. As a powerful computational tool, Markov chain Monte Carlo (MCMC) methods are widely employed for posterior distribution numerical simulation [8-15].

The MCMC method generates samples as the states of a special Markov chain whose stationary distribution is equal to the posterior distribution. Under the assumption of ergodicity, a sufficient number of posterior model parameter samples can be obtained from the Markov chain simulation. Many classical MCMC algorithms have been developed based on the theory of the Metropolis–Hastings (MH) algorithm. Two examples are the transitional Markov chain Monte Carlo (TMCMC) algorithm [12,14-15], which can efficiently deal with the large difference between the prior and posterior distributions in practical problems, and the Hamiltonian Monte Carlo (HMC) method [8,16-17] which enables fast exploration of the state space by simply introducing auxiliary

"momentum" variables and replacing the simple random-walk proposal candidates used in MH algorithms with Hamiltonian dynamics proposals. However, these generic MCMC methods are not tailored to any specific Bayesian updating problem. Further extension and optimization of generic MCMC methods [18-20] are useful for exploring their potential in addressing specific problems, but their effectiveness is still correlated with and limited by their flexibility. However, the more algorithmic parameters that need to be tuned, the greater the difficulty for practitioners.

In recent years, the application of neural networks has made the extension and optimization of generic MCMC methods easier and tractable. A large number of MCMC algorithms, enhanced by neural networks, have been proposed [21-23]. These algorithms employ neural networks to replace and enhance specific components within MCMC algorithms to get better proposals for targeted problems, with the goal of achieving fast convergence and efficient exploration of the sampling processes. With the improvement of the performance for specific problems after training, the generalization ability of the methods usually decreases. As a consequent drawback, the embedded neural networks often need to be retrained when faced with new tasks, which is time-consuming and significantly weakens the competitiveness of the methods.

Meta-learning techniques [24,25] efficiently acquire the common knowledge contained in a set of similar tasks through various ingenious designs of learning algorithms, aiming to guarantee the generalization ability of learners, thereby enabling them to tackle a range of similar tasks by learning from only one or a few of them. Based on the idea of meta-learning, once trained, the MCMC algorithms can be employed directly for a bunch of similar Bayesian updating tasks, thereby significantly reducing training time and greatly enhancing the competitiveness of the neural network-based algorithms. This paper achieves this goal by combining two strategies [26,27]:

theoretical improvement and network training.

A neural network enhanced SGHMC (NN-SGHMC) algorithm [23] selects key potential adjustable parts of the stochastic gradient Hamiltonian Monte Carlo (SGHMC) algorithm and replaces them with neural networks for training; its meta-learning potential is shown for some Bayesian deep learning tasks. However, its generalization ability is limited, and it cannot be employed directly for structural dynamic model updating due to the variety and complexity of dynamic models.

This paper presents a newly developed adaptive meta-learning stochastic gradient Hamiltonian Monte Carlo (AM-SGHMC) algorithm. In AM-SGHMC, the replaced key adjustable parts are based on NN-SGHMC, but a newly developed adaptive meta-learning technique is introduced to guarantee the scale-invariance of the trained sampler with respect to the posterior PDF, which greatly improves its generalization ability. In addition, the loss function and its back-propagation path are carefully designed, making them more suitable for network training in a Markov chain environment. As for model updating problems, a prior boundary treatment approach is proposed, which enables the free exploration in the early stage of network training. The above efforts successfully enable meta-learning for Bayesian updating of structural dynamic models.

The remainder of this paper is organized as follows. In Section 2, the main challenges and a blueprint of the AM-SGHMC are briefly introduced. In Section 3, the main contributions of the AM-SGHMC are further detailed. In Section 4, the AM-SGHMC algorithm is summarized in a pseudocode. In Section 5, two examples involving Bayesian updating of a multi-story shear-building model and a multi-story braced-frame building model, are used to demonstrate the effectiveness and generalization ability of the proposed method. Finally, we conclude in Section 6 with a summary of this study.

## 2. AM-SGHMC: ADAPTIVE META-LEARNING STOCHASTIC GRADIENT HAMILTONIAN MONTE CARLO

### *2.1 Problem definition and framework for SG-MCMC*

AM-SGHMC is based on a *complete* framework for stochastic gradient MCMC (SG-MCMC) [28]. Here, "*complete*" signifies that the SG-MCMC framework is capable of accommodating any continuous Markov process that is designed to generate samples from a target distribution. In the sample space $\boldsymbol{\theta} \in \mathbb{R}^D$, SG-MCMC treats the samples as the positions of particles and performs stochastic dynamic simulation of an Itô diffusion process to generate Markov chains. For the Itô diffusion process, the stationary distribution of particles in a potential energy field $U(\boldsymbol{\theta})$ is given by $\pi(\boldsymbol{\theta}) \propto \exp(-U(\boldsymbol{\theta}))$.

In the context of the Bayesian approach, the stationary distribution should correspond to the posterior probability density function (PDF), i.e., $\pi(\boldsymbol{\theta}) = p(\boldsymbol{\theta}|\mathcal{D})$, where the posterior PDF is derived from Bayes' theorem:

$$p(\boldsymbol{\theta}|\mathcal{D}) = p(\mathcal{D}|\boldsymbol{\theta})p(\boldsymbol{\theta})/p(\mathcal{D}) \propto p(\mathcal{D}|\boldsymbol{\theta})p(\boldsymbol{\theta}) \qquad (1)$$

Here, $\boldsymbol{\theta}$ represents the model parameter vector, $p(\boldsymbol{\theta})$ is the prior PDF, and $p(\mathcal{D}|\boldsymbol{\theta})$ is the likelihood function, which quantifies the probability of observing the data $\mathcal{D}$ given different values of the model parameters $\boldsymbol{\theta}$. The evidence function $p(\mathcal{D}) = \int p(\mathcal{D}|\boldsymbol{\theta})p(\boldsymbol{\theta})\, d\boldsymbol{\theta}$ is a normalizing constant in the denominator that ensures the posterior PDF integrates to one.

Consequently, the potential energy field can be set as:

$$U(\boldsymbol{\theta}) = -\log(p(\boldsymbol{\theta}|\mathcal{D})) + c^{**} = -\log(p(\mathcal{D}|\boldsymbol{\theta})) - \log(p(\boldsymbol{\theta})) + c^* \qquad (2)$$

where $c^*$ can be set as an arbitrary constant to facilitate calculation, thereby $c^{**} = -\log(p(\mathcal{D})) + c^*$. It is worth noting that, as with other Monte Carlo methods, the evidence function $p(\mathcal{D})$ in Equation (1), which contains an intractable multidimensional integral, is not involved here.

In SG-MCMC, the augmented state space is defined as $\mathbf{z} = (\boldsymbol{\theta}, \boldsymbol{p})$, where $\boldsymbol{p}$ denotes a set of auxiliary variables, and the Hamiltonian is defined as $H(\mathbf{z}) = U(\boldsymbol{\theta}) + \mathcal{g}(\boldsymbol{p}|\boldsymbol{\theta})$, based on the designed PDF of auxiliary variables $\pi(\boldsymbol{p}|\boldsymbol{\theta}) \propto \exp(-\mathcal{g}(\boldsymbol{p}|\boldsymbol{\theta}))$, such that the augmented stationary distribution is given by $\pi(\mathbf{z}) \propto \exp(-H(\mathbf{z}))$.

The dynamic simulation within SG-MCMC is founded on an Itô diffusion process, which can be expressed by the continuous-time stochastic differential equation (SDE):

$$d\mathbf{z} = \boldsymbol{f}(\mathbf{z})dt + \sqrt{2\boldsymbol{D}(\mathbf{z})}d\boldsymbol{W}(t) \qquad (3)$$

where $\boldsymbol{f}(\mathbf{z})$, $\boldsymbol{W}(t)$ and $\boldsymbol{D}(\mathbf{z})$ are the deterministic drift, Wiener process and diffusion matrix, respectively.

The drift $\boldsymbol{f}(\mathbf{z})$ can be explicitly parameterized as:

$$\boldsymbol{f}(\mathbf{z}) = -[\boldsymbol{D}(\mathbf{z}) + \boldsymbol{Q}(\mathbf{z})]\nabla_\mathbf{z} H(\mathbf{z}) + \boldsymbol{\Gamma}(\mathbf{z}) \qquad (4)$$

$$\boldsymbol{\Gamma}_i(\mathbf{z}) = \sum_{j=1}^{d} \frac{\partial}{\partial z_j}[\boldsymbol{D}_{ij}(\mathbf{z}) + \boldsymbol{Q}_{ij}(\mathbf{z})] \qquad (5)$$

where $\boldsymbol{Q}(\mathbf{z})$ and $\boldsymbol{\Gamma}(\mathbf{z})$ represent the curl matrix and a correction term, respectively. For brevity, we define $\overleftarrow{\boldsymbol{A}(\mathbf{z}) \cdot \nabla_\mathbf{z}} \triangleq \left(\nabla_\mathbf{z} \cdot \boldsymbol{A}^\mathrm{T}(\mathbf{z})\right)^\mathrm{T}$ for any vector $\mathbf{z}$ and matrix $\boldsymbol{A}(\mathbf{z})$, such that Equation (5) can be represented efficiently by a single matrix equation $\boldsymbol{\Gamma}(\mathbf{z}) = \overleftarrow{[\boldsymbol{D}(\mathbf{z}) + \boldsymbol{Q}(\mathbf{z})] \cdot \nabla_\mathbf{z}}$.

The *completeness* of the above framework has been proved by [28]:

(1) $\pi(\mathbf{z}) \propto \exp(-H(\mathbf{z}))$ is a stationary distribution of the SDEs (3)-(5) for any pair of skew-symmetric matrix $\boldsymbol{Q}(\mathbf{z})$ and positive semi-definite matrix $\boldsymbol{D}(\mathbf{z})$;

(2) for any Itô diffusion process that possesses a unique stationary distribution $\pi(\mathbf{z})$, under mild conditions there exist matrices $\boldsymbol{Q}(\mathbf{z})$ and $\boldsymbol{D}(\mathbf{z})$ such that the process is governed by SDEs (3)-(5).

The completeness of this framework for SG-MCMC and the flexibility in choosing $Q(z)$ and $D(z)$ are very suitable for the embedding of neural networks.

## 2.2 Neural network enhancement of SGHMC and its challenges

SG-MCMC is a framework that encompasses all sampling methods with continuous Markov processes, including SGHMC, the predecessor of our AM-SGHMC, based on specific choices of $H(z)$, $Q(z)$ and $D(z)$.

During the stochastic dynamic simulation of SGHMC, $\boldsymbol{\theta}$ and the auxiliary vector $\boldsymbol{p}$ are regarded as the position and momentum of the particles respectively. Assuming an identity mass matrix $\boldsymbol{M} = \mathbf{I}_D$, the Hamiltonian is defined as $H(z) = U(\boldsymbol{\theta}) + \frac{1}{2}\boldsymbol{p}^\mathrm{T}\boldsymbol{p}$, where $U(\boldsymbol{\theta})$ and $\mathcal{G}(\boldsymbol{p}) = \frac{1}{2}\boldsymbol{p}^\mathrm{T}\boldsymbol{p}$ can be regarded as potential and kinetic energy terms, respectively. Consequently, $\boldsymbol{p}$ has the same dimension as $\boldsymbol{\theta}$, and in the stationary distribution, it follows the standard Gaussian distribution $\boldsymbol{p} \sim \mathcal{N}(\mathbf{0}_D, \mathbf{I}_D)$.

In the SG-MCMC framework, the curl matrix $Q(z)$ and diffusion matrix $D(z)$ for SGHMC are chosen as:

$$Q(z) = \begin{bmatrix} 0 & -G \\ G & 0 \end{bmatrix} \tag{6}$$

$$D(z) = \begin{bmatrix} 0 & 0 \\ 0 & C \end{bmatrix} \tag{7}$$

where $C$ is a positive definite matrix called here the damping matrix and $G$ is called here the gyroscopic-coupling matrix. According to the chosen $H(z)$, $Q(z)$ and $D(z)$, the continuous-time SDEs governing the dynamics of SGHMC can be derived as:

$$d\boldsymbol{\theta} = \boldsymbol{G}\boldsymbol{p}\,dt \tag{8}$$

$$d\boldsymbol{p} = -\boldsymbol{C}\boldsymbol{p}\,dt - \boldsymbol{G}\nabla_{\boldsymbol{\theta}}U(\boldsymbol{\theta})dt + \sqrt{2\boldsymbol{C}}d\boldsymbol{W}(t) \tag{9}$$

where the potential energy gradient $\nabla_{\boldsymbol{\theta}}U(\boldsymbol{\theta})$ can be calculated using the auto-differentiation technique [8]. Then, the corresponding discretized update rule with modified forward Euler discretization of SGHMC with step-size $\eta$ is:

$$\boldsymbol{p}_{t+1} = (1 - \eta \boldsymbol{C})\boldsymbol{p}_t - \eta \boldsymbol{G}\nabla_{\boldsymbol{\theta}_t}U(\boldsymbol{\theta}_t) + \boldsymbol{\epsilon}_t, \qquad \boldsymbol{\epsilon}_t \sim \mathcal{N}(\boldsymbol{0}_D, 2\eta \boldsymbol{C}) \qquad (10)$$

$$\boldsymbol{\theta}_{t+1} = \boldsymbol{\theta}_t + \eta \boldsymbol{G}\boldsymbol{p}_{t+1} \qquad (11)$$

It can be seen that, in SGHMC, the gyroscopic-coupling matrix $\boldsymbol{G}$ in $\boldsymbol{Q}(\boldsymbol{z})$ mainly controls the interaction between $\boldsymbol{\theta}$ and $\boldsymbol{p}$, while the role of the damping matrix $\boldsymbol{C}$ in $\boldsymbol{D}(\boldsymbol{z})$ is to control the intensity of damping to counteract the influence of the corresponding intensity of the random impulse $\boldsymbol{\epsilon}_t$. In standard SGHMC, $\boldsymbol{G}$ is usually chosen as the identity matrix $\mathbf{I}_D$ and $\boldsymbol{C}$ is a constant matrix, which can be chosen as in [20], where other choices of the mass matrix than $\boldsymbol{M} = \mathbf{I}_D$ are also examined.

Rather than treating the matrices $\boldsymbol{G}$ and $\boldsymbol{C}$ as constants, Gong et. al. [23] point out that making them functions of $\boldsymbol{z}$ can lead to significant improvements in mixing as well as reduction of sample bias. They use two neural networks to learn functions for diagonal versions of the matrices $\boldsymbol{G}(\boldsymbol{z})$ and $\boldsymbol{C}(\boldsymbol{z})$ using as inputs the potential energy $U(\boldsymbol{\theta})$, the potential energy gradient $\partial_{\boldsymbol{\theta}}U(\boldsymbol{\theta})$, and the momentum $\boldsymbol{p}$. This choice of inputs actually ensures the translation-invariance of the sampling process to the posterior PDF by avoiding the direct use of $\boldsymbol{z} = (\boldsymbol{\theta}, \boldsymbol{p})$, which is a key factor contributing to the generalization ability.

However, there are still some challenges in achieving meta-learning for Bayesian updating of structural dynamic models. It is found that, even for data from the same structure, both the arbitrariness of constant $c^*$ in Equation (2) and the large variation in the scale of the posterior PDF will lead to a large variation in the network inputs, while the scale variation also demands a large variation in the network outputs. This phenomenon indicates a large difference in the input-output relationship between different tasks, which means that the trained neural networks are no longer suitable for many new tasks. Therefore, a new network embedding design ensuring the scale-invariance of the sampling process with respect to the posterior PDF is desired.

In terms of network training, for Bayesian updating problems of structural dynamic models, the potential energy $U(\boldsymbol{\theta})$ changes more dramatically, necessitating more reliable training techniques. There are two main aspects that can be improved. Firstly, sample-based loss functions that disregard sample order are often selected to update network parameters through direct backpropagation. Without making full use of the property of Markov chains, both the naive loss function and its straightforward backpropagation path introduce some interference into the network training. Secondly, bounded priors, which constrain parameters to a reasonable range, are rarely employed in gradient-based MCMC methods due to the singularity of their gradients at the boundary. However, bounded priors are crucial for ensuring full exploration of the state space during training by avoiding the complex rare event regions.

Our contributions to addressing these challenges are detailed in Section 3.

## *2.3 Blueprint of AM-SGHMC*

Before going into details, a higher-level blueprint is first given in this section for our proposed algorithm. In AM-SGHMC, two neural networks are proposed to substitute and optimize the matrices $\boldsymbol{Q}(\boldsymbol{z})$ and $\boldsymbol{D}(\boldsymbol{z})$ for specific sampling problems.

Considering the feasibility of network training, the parameterization of matrices $\boldsymbol{Q}(\boldsymbol{z})$ and $\boldsymbol{D}(\boldsymbol{z})$ in SG-MCMC are designed as follows:

$$\boldsymbol{Q}(\boldsymbol{z}) = \begin{bmatrix} \boldsymbol{0} & -\boldsymbol{G}(\boldsymbol{z}) \\ \boldsymbol{G}(\boldsymbol{z}) & \boldsymbol{0} \end{bmatrix} \tag{12}$$

$$\boldsymbol{D}(\boldsymbol{z}) = \begin{bmatrix} \boldsymbol{0} & \boldsymbol{0} \\ \boldsymbol{0} & \boldsymbol{C}(\boldsymbol{z}) \end{bmatrix} \tag{13}$$

where

$$\boldsymbol{G}(\boldsymbol{z}) = \operatorname{diag}\left[\boldsymbol{f}_{\phi_Q}(\boldsymbol{z})\right] \tag{14}$$

$$C(z) = \text{diag}[f_{\phi_D}(z)] \tag{15}$$

where each element is:

$$f_{\phi_Q,i}(z) = \sigma_i \cdot \left(c_1 + f_{\phi_Q}(\widehat{U}(\boldsymbol{\theta}), p_i, Cate_i)\right) \tag{16}$$

$$f_{\phi_D,i}(z) = c_2 + f_{\phi_D}(\widehat{U}(\boldsymbol{\theta}), p_i, \partial_{\theta_i}\widehat{U}^*(\boldsymbol{\theta}), Cate_i) \tag{17}$$

where $f_{\phi_Q} > 0$ and $f_{\phi_D} > 0$ are the outputs of two embedded neural networks, and $c_1$ and $c_2$ are two small positive constants to prevent the vanishing of matrices $Q(z)$ and $D(z)$. In order to maintain scale-invariance with respect to the posterior PDF, which enhances meta-learning ability, adaptive scale estimators $\sigma_i$ and adaptively normalized network inputs $\widehat{U}(\boldsymbol{\theta})$ and $\partial_{\theta_i}\widehat{U}^*(\boldsymbol{\theta})$ are adopted, as detailed in Section 3.1, based on their raw materials, $\theta_i$, $U(\boldsymbol{\theta})$ and $\partial_{\theta_i}U(\boldsymbol{\theta})$, which can be obtained during the sampling process. The $\theta_i$ and $p_i$ appearing in the network inputs are the $i$-th components of $\boldsymbol{\theta}$ and $\boldsymbol{p}$, respectively, corresponding to the $i$-th model parameter. Finally, the network input $Cate_i$ is an encoded category of the $i$-th parameter to allow different sampling strategies for different categories of parameters, which can also be viewed as a constant function of the state space $z = (\boldsymbol{\theta}, \boldsymbol{p})$. Thus, the inputs of the networks, $\widehat{U}(\boldsymbol{\theta})$, $p_i$, $\partial_{\theta_i}\widehat{U}^*(\boldsymbol{\theta})$ and $Cate_i$, are all functions of state space $z$, which results in all elements in matrices $Q(z)$ and $D(z)$ being functions of $z$. More computational details and explanations about this adaptive input/output processing are shown in Section 3.1.

Based on SDEs (3)-(5) with modified forward Euler discretization, the sampling process using discretized dynamics with step-size $\eta$ is:

$$\begin{aligned}\boldsymbol{p}_{t+1} = &(1 - \eta C(z_t))\boldsymbol{p}_t - \eta G(z_t)\nabla_{\theta_t} U(\boldsymbol{\theta}_t) \\ &+ \eta \left(\overleftarrow{G(z_t) \cdot \nabla_{\theta_t}} + \overleftarrow{C(z_t) \cdot \nabla_{p_t}}\right) + \epsilon_t\end{aligned} \tag{18}$$

$$\boldsymbol{\theta}_{t+1} = \boldsymbol{\theta}_t + \eta \boldsymbol{G}(\hat{\boldsymbol{z}}_t)\boldsymbol{p}_{t+1} - \eta \overleftarrow{\boldsymbol{G}(\hat{\boldsymbol{z}}_t) \cdot \nabla_{\boldsymbol{p}_{t+1}}} \qquad (19)$$

where $t$ is the discretized time; $\boldsymbol{z}_t$, $\hat{\boldsymbol{z}}_t$ and $\boldsymbol{\epsilon}_t$ are defined as $\boldsymbol{z}_t = (\boldsymbol{\theta}_t, \boldsymbol{p}_t)$, $\hat{\boldsymbol{z}}_t = (\boldsymbol{\theta}_t, \boldsymbol{p}_{t+1})$ and $\boldsymbol{\epsilon}_t \sim \mathcal{N}(\boldsymbol{0}, 2\eta \boldsymbol{C}(\boldsymbol{z}_t))$, respectively.

Whenever the initial state $\boldsymbol{z}_0 = (\boldsymbol{\theta}_0, \boldsymbol{p}_0)$ is determined, parameter samples from a Markov chain can be generated by performing Equations (18)-(19) iteratively. After simulating $K$ parallel Markov chains for $T$ steps, a set of samples $\boldsymbol{\Theta}_{T,\tau}^K = \{\{\boldsymbol{\theta}_{s\tau}^k\}_{s=1}^{\left[\frac{T}{\tau}\right]}\}_{k=1}^{K}$ with sampling interval $\tau$ can be obtained. The PDF of the samples $\boldsymbol{\Theta}_{T,\tau}^K$ is written as $q(\boldsymbol{\theta}|\mathcal{D})$, which can be estimated as $\bar{q}(\boldsymbol{\theta}|\mathcal{D}) = \frac{1}{K}\frac{1}{\left[\frac{T}{\tau}\right]}\sum_{k=1}^{K}\sum_{s=1}^{\left[\frac{T}{\tau}\right]}\phi\left(\boldsymbol{\theta}; \boldsymbol{\theta}_{s\tau}^k, c_{op}\boldsymbol{\Sigma}_{\boldsymbol{\Theta}_{T,\tau}^K}\right)$ using the samples $\boldsymbol{\Theta}_{T,\tau}^K$, where $\phi(\cdot; \boldsymbol{\mu}, \boldsymbol{\Sigma})$ is a multidimensional Gaussian PDF with mean $\boldsymbol{\mu}$ and covariance matrix $\boldsymbol{\Sigma}$, $\boldsymbol{\Sigma}_{\boldsymbol{\Theta}_{T,\tau}^K}$ is the covariance matrix of samples $\boldsymbol{\Theta}_{T,\tau}^K$, and $c_{op}$ is set to minimize $\frac{1}{K}\frac{1}{\left[\frac{T}{\tau}\right]}\sum_{k=1}^{K}\sum_{s=1}^{\left[\frac{T}{\tau}\right]}\log \bar{q}(\boldsymbol{\theta}_{s\tau}^k|\mathcal{D})$. Other more accurate kernel density functions, such as that detailed in [10], can also be used.

The loss function of AM-SGHMC algorithm can be approximated as:

$$Loss = -\mathcal{L}_{VI}(\boldsymbol{\Theta}_{T,\tau}^K) = \frac{1}{K}\frac{1}{\left[\frac{T}{\tau}\right]}\sum_{k=1}^{K}\sum_{s=1}^{\left[\frac{T}{\tau}\right]}[U(\boldsymbol{\theta}_{s\tau}^k) + \log \bar{q}(\boldsymbol{\theta}_{s\tau}^k|\mathcal{D})] \qquad (20)$$

which is a Monte Carlo estimate of the negative Evidence Lower Bound (ELBO), $\mathcal{L}_{VI}(q) = -\mathbb{E}_q[U(\boldsymbol{\theta})] + \mathbb{H}[q]$, consisting of an energy term $-\mathbb{E}_q[U(\boldsymbol{\theta})]$ and an information entropy term $\mathbb{H}[q]$. While maximizing the variational lower-bound ELBO $\mathcal{L}_{VI}(q)$, reducing the term $Loss = -\mathcal{L}_{VI}(q)$ is like minimizing the KL-divergence $\mathrm{KL}[q \parallel \pi]$, which encourages the sampler to generate samples that converge faster to the stationary distribution $\pi(\boldsymbol{\theta})$ in a finite number of steps $T$. During network training, only the gradients of $\mathbb{H}[q]$ are needed, so that the *Stein gradient estimator* detailed in [23] is used to directly estimate each gradient $\nabla_{\boldsymbol{\theta}_{s\tau}^k} \log q(\boldsymbol{\theta}_{s\tau}^k|\mathcal{D})$ using the samples $\boldsymbol{\Theta}_{T,\tau}^K$, which is more effective than first approximating $\bar{q}(\boldsymbol{\theta}|\mathcal{D})$ and then applying the auto-differentiation technique. The specific form of the loss function and more computational details are shown in Section 3.2.

For a specific sampling problem, by training with the loss function similar to Equation (20) through an efficient back-propagation path detailed in a later section, AM-SGHMC sampler can reliably achieve a good performance with considerable generalization ability for a set of homogeneous sampling problems. Its meta-learning ability is demonstrated in two examples of Bayesian updating of structural dynamic models in Section 5.

An additional issue is that, how to rapidly assess whether the trained sampler can achieve satisfactory generalization performance, when faced with a new and significantly different task. Since the inputs to the networks consist solely of local component-wise information, the conditional posterior PDFs for each parameter, which more intuitively reflect the local component-wise information, are crucial in influencing the performance of the trained sampler. Due to the scale-invariance of the sampler with respect to the posterior PDF, one can briefly assess whether the shapes of the conditional posterior PDFs for the new task resemble those of parameters belonging to the same categories in the training task. To clearly observe the tail trends of the conditional distributions, we recommend visualizing the cross-section of the potential energy function, which is the negative logarithm of the conditional posterior PDF plus an unknown constant, as shown in Equation (2). If the shapes resemble each other, it can be initially concluded that the trained sampler can achieve satisfactory generalization performance for this new task.

## 3. MAIN CONTRIBUTIONS IN AM-SGHMC

### *3.1 Adaptive input/output processing with scale-invariance for the posterior PDF*

In the referenced NN-SGHMC algorithm, the neural networks are directly used to optimize the matrices $Q(z)$ and $D(z)$, that is, each element in the matrices $G(z)$ and $C(z)$ is calculated as:

$$f_{\phi_Q,i}(z) = c_1 + f_{\phi_Q}(U(\boldsymbol{\theta}), p_i) \tag{21}$$

$$f_{\phi_D,i}(z) = c_2 + f_{\phi_D}\left(U(\boldsymbol{\theta}), p_i, \partial_{\theta_i}U(\boldsymbol{\theta})\right) \tag{22}$$

By training with a loss function similar to Equation (20), an NN-SGHMC sampler can achieve a satisfactory performance for specific sampling problems. However, given the arbitrariness of the potential energy reference point and the large variation in the scale of the posterior PDF, the generalization ability of the NN-SGHMC algorithm is limited, preventing it from achieving meta-learning for Bayesian updating of structural dynamic models. Therefore, an input/output processing procedure with scale-invariance with respect to the posterior PDF is developed and detailed in this section.

As shown in Equation (2), the constant $c^*$ is of arbitrary value and so the potential energy input $U(\boldsymbol{\theta})$ can be very different even in the same model updating problem. To fix this issue, the normalization shown in Equation (23) is applied to the potential energy input $U(\boldsymbol{\theta})$.

$$\widehat{U}(\boldsymbol{\theta}) = \frac{U(\boldsymbol{\theta}) - \mu_U}{\sqrt{2D}\sigma_U} \tag{23}$$

where $\widehat{U}(\boldsymbol{\theta})$ is the normalized potential energy, $D$ is the dimension of the parameter space, and the lower-order moments $\mu_U$ and $\sigma_U$ are the mean and standard deviation of the potential energy $U(\boldsymbol{\theta})$, respectively, which can be estimated adaptively during burn-in similar to those adopted in Adam [29]. The numerator part eliminates the effect of the constant c, while making the $\widehat{U}(\boldsymbol{\theta})$ scale-invariant with respect to the posterior PDF. As for the denominator part, it can make the minimum value of $\widehat{U}(\boldsymbol{\theta})$ as stable as possible when the dimension $D$ of the parameter space changes from task to task.

In addition to the normalization of the potential energy input, the following input/output processes are performed to achieve scale-invariance with respect to the posterior PDF, based on the adaptively estimated standard deviation $\sigma_i, i = 1, \cdots, D$ of each dimension of the parameter posterior PDF during burn-in. The potential energy gradient input $\partial_{\theta_i} U(\boldsymbol{\theta})$ in Equation (22) is changed into:

$$\partial_{\theta_i}\widehat{U}^*(\boldsymbol{\theta}) = \sigma_i \cdot \partial_{\theta_i}\widehat{U}(\boldsymbol{\theta}) \tag{24}$$

And the output for elements of matrix $\boldsymbol{Q}(\boldsymbol{z})$ shown in Equation (21) is changed into:

$$\boldsymbol{f}_{\phi_Q,i}(\boldsymbol{z}) = \sigma_i \cdot \left(c_1 + f_{\phi_Q}(\widehat{U}(\boldsymbol{\theta}), p_i)\right) \tag{25}$$

The above input/output processing approach is proposed based on the phenomenon that the shape of the conditional posterior PDF for parameters of the same category is relatively similar. Considering that the difference in PDF shape between different parameter categories may be relatively large, the parameter category is also encoded as one of the neural network inputs to accommodate different sampling strategies for different shapes. Specifically, the categories can be divided based on the functionality of the parameters, and subsequently, one-hot encoding, which is one of the most important encoding techniques for categorical data, is recommended. Of course, other ad-hoc encoding techniques can also be employed. Denoting the encoded category of the $i$-th parameter as $Cate_i$, the Equations (25) and (22) can be finally updated into Equations (16)-(17) shown in Section 2.3.

The process of adaptive estimates and a simple proof of the scale-invariance are presented in Appendix A and Appendix B, respectively.

With the computational details in Equations (23)-(24) as well as Appendix A, the newly developed adaptive meta-learning procedure, AM-SGHMC, is formed as shown in Figure 1, which aims to improve the generalization ability of the neural network-based sampler for Bayesian updating problems of structural dynamic models.

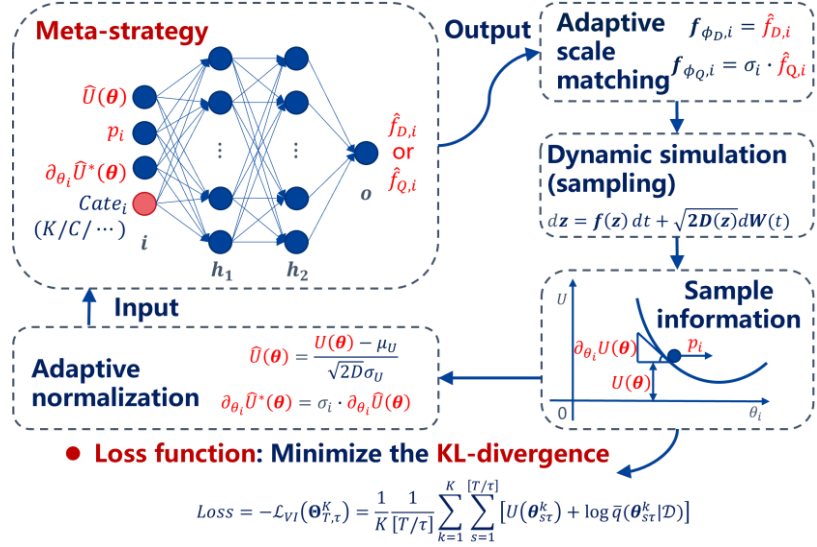

**Figure 1.** Schematic of AM-SGHMC.

## *3.2 Loss function and its back-propagation path adapted to the Markov chain environment*

As for the back-propagation path during training, truncated back-propagate through time (BPTT) is applied in NN-SGHMC algorithm. That is, manually stop the gradient flow through the input of $Q$ and $D$ matrices every $T_{TBPTT}$ steps to avoid computing higher order gradients, where the truncated step $T_{TBPTT}$ is typically 10 to 20. In addition, the sampling interval for training is typically $\tau = 1$.

The schematic of the forward calculation process involved in one back-propagation of NN-SGHMC algorithm is shown in Figure 2. During back-propagation, the gradient flow will accumulate from the loss function against the arrow direction to the neural network parameters $\phi$, guiding their update. Specifically, the loss function will firstly guide the optimization of the samples, and then the optimization of each sample will guide the optimization of the previous sample and the corresponding matrices $G(z)$ and $C(z)$ based on the dynamic simulation process, Equations (18)-(19), up to the initial sample. Finally, the optimization of all the matrices $G(z_{t_0+s}^k)$, $C(z_{t_0+s}^k)$ and $G(\hat{z}_{t_0+s}^k)$, $s = 0, \cdots, T_{TBPTT}$, will jointly guide the optimization of the network parameters $\phi$.

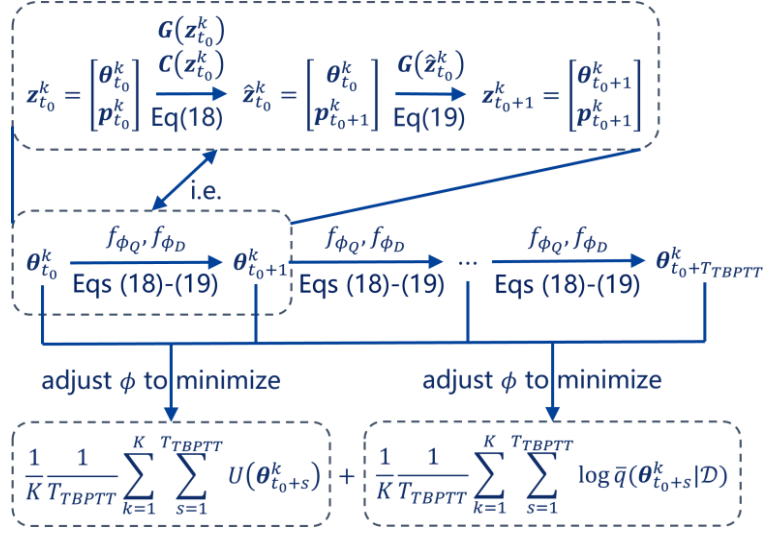

**Figure 2.** Schematic of the forward calculation process involved in one back-propagation of NN-SGHMC algorithm.

Since the samples generated by MCMC methods are the states of a Markov chain, the generation of each sample can be fully determined by the previous adjacent sample only, which means that the information of all the other samples will be redundant when the previous adjacent sample is known. Accordingly, in AM-SGHMC algorithms, each sample $\theta_{t_0+s^*}^k$ is determined only by the matrices $G(z_{t_0+s^*-1}^k)$, $C(z_{t_0+s^*-1}^k)$ and $G(\hat{z}_{t_0+s^*-1}^k)$ output from the networks in the last step when the previous adjacent sample $\theta_{t_0+s^*-1}^k$ is known, i.e., each matrix $G(z)$ or $C(z)$ is responsible for only one subsequent sample. However, according to the back-propagation path in Figure 2, each matrix $G(z)$ or $C(z)$ is responsible for all subsequent samples within the truncated step $T_{TBPTT}$, which leads to large noise and affects the training effect.

Thus, we firstly improve the back-propagation path by manually stop the gradient flow through the states at each step, which is similar to setting $T_{TBPTT} = 1$, to ensure the stability of the training process. To distinguish, the number of steps involved in each back-propagation of AM-SGHMC is denoted as $T_T$ rather than $T_{TBPTT}$. This improvement can be achieved directly for the energy term of loss function on the left of Figure 2, but for the information entropy term on the right, the estimate of the sample distribution $q(\boldsymbol{\theta}|\mathcal{D})$ needs to be considered.

When estimating $\bar{q}(\boldsymbol{\theta}^k_{t_0+s^*}|\mathcal{D})$ to guide the sampling of $\boldsymbol{\theta}^k_{t_0+s^*}$, considering that the samples after $t_0 + s^*$ are affected by $\boldsymbol{\theta}^k_{t_0+s^*}$ and have considerable uncertainty because they have not yet been generated when sampling $\boldsymbol{\theta}^k_{t_0+s^*}$, we only utilize the previous samples, i.e., $\{\{\boldsymbol{\theta}^k_{t_0+s}\}_{s=0}^{s^*}\}_{k=1}^{K}$, rather than all the samples $\{\{\boldsymbol{\theta}^k_{t_0+s}\}_{s=0}^{T_T}\}_{k=1}^{K}$. This change makes the order of samples reflected in the loss function. We denote the sample distribution estimated with samples $\{\{\boldsymbol{\theta}^k_{t_0+s}\}_{s=0}^{s^*}\}_{k=1}^{K}$ as $\bar{q}_{s^*}(\boldsymbol{\theta}|\mathcal{D})$. Although the repeated estimation of $\bar{q}_{s^*}(\boldsymbol{\theta}|\mathcal{D})$ with different $s^*$ takes more time, it is insignificant compared to the calculation of the energy term.

Finally, considering that the estimated distribution is not accurate when the number of samples is small, the first $M$ estimates $\{\bar{q}_s(\boldsymbol{\theta}^k_{t_0+s}|\mathcal{D})\}_{s=1}^{M}$ are not used in the estimation of the information entropy term. So that the loss function for each back-propagation of AM-SGHMC algorithm can be written more concretely as:

$$Loss_{t_0} = \frac{1}{K}\frac{1}{\left[\frac{T_T}{\tau}\right]}\sum_{k=1}^{K}\sum_{s=1}^{\left[\frac{T_T}{\tau}\right]}U(\boldsymbol{\theta}^k_{t_0+s\tau}) + \frac{1}{K}\frac{1}{\left[\frac{T_T}{\tau}\right]-M}\sum_{k=1}^{K}\sum_{s=M+1}^{\left[\frac{T_T}{\tau}\right]}\log \bar{q}_s(\boldsymbol{\theta}^k_{t_0+s\tau}|\mathcal{D}) \quad (26)$$

where $\tau = 1$ is recommended.

In addition, once $T_T$ is too large, the framework for SG-MCMC already ensures that the long-term distribution of the samples is consistent with the target posterior PDF. This means that the loss function has already been minimized and is not suitable for further optimizing the sampling strategy networks. Thus, choosing $T_T$ in the range of 10 to 20, as in NN-SGHMC, remains a good choice. In our experiments below, for example, we set $T_T = 15$ and $M = 3$.

To sum up, the schematic of the forward calculation process involved in one back-propagation of AM-SGHMC algorithm is shown in Figure 3, where the dashed lines indicate the stop gradient operations.

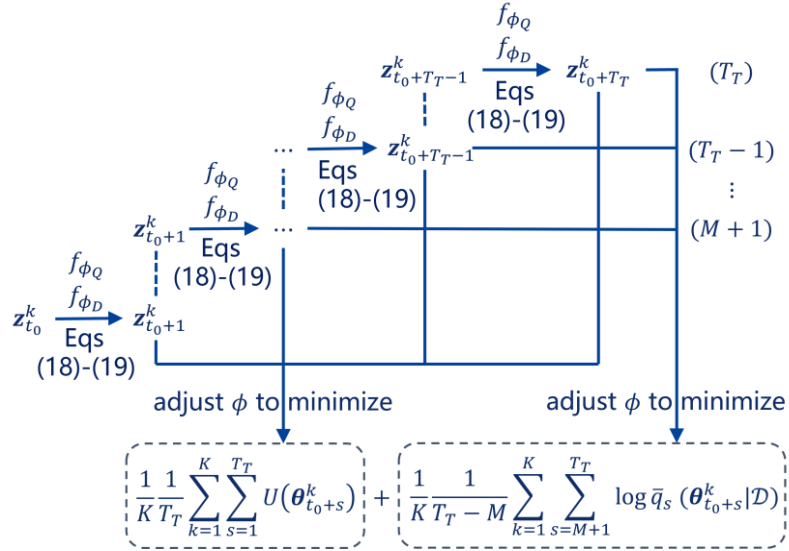

**Figure 3.** Schematic of the forward calculation process involved in one back-propagation of AM-SGHMC algorithm.

### *3.3 Treatment of prior boundaries enables adequate training*

In addition to the development of the AM-SGHMC algorithm itself, the treatment of PDFs defined by model updating problems is also crucial to ensure the well training of the neural network. A treatment of commonly used bounded priors is proposed in this section, which allows it to be used in MCMC methods with dynamics proposals.

Bounded priors, such as uniform distributions and truncated Gaussian distributions,

can limit the parameters to a reasonable range, so as to reduce unnecessary exploration of the state space and avoid the failure of the MCMC method due to the complexity of the low probability regions.

For the dynamics proposals utilizing the potential energy gradient, ideally the sample moving to the boundary will be bounced due to the infinite gradient at the boundary. However, due to the discretization of dynamics during calculation, the samples will hardly be exactly at the boundary, so that the infinite gradient at the boundary cannot be sensed.

In general, the model parameter vector $\boldsymbol{w}$ is equivalent to the state $\boldsymbol{\theta}$ of the Markov chain, but we change this when there is a bounded prior. In order to enable the application of bounded priors in MCMC methods with dynamics proposals, a stochastic variable transformation is proposed, which can extend the parameter space near the boundary to infinity, thus forming a new state space suitable for dynamics proposals.

Specifically, when encountering a bounded prior, for $i$-th parameter $w_i$ bounded by interval $[b_{i,1} - \delta_{i,1}, b_{i,2} + \delta_{i,2}]$ where $b_{i,1} \leq b_{i,2}$, denoting its corresponding state as $\theta_i \in \mathbb{R}$, we define a monotonic increasing stochastic variable transformation from state space into parameter space as follows:

$$w_i = \begin{cases} f(\theta_i; b_{i,1}, \delta_{i,1}), & \theta_i < b_{i,1} \\ \theta_i, & b_{i,1} \leq \theta_i \leq b_{i,2} \\ f(\theta_i; b_{i,2}, \delta_{i,2}), & \theta_i > b_{i,2} \end{cases} \quad (27)$$

where the function $f(\theta; b, \delta)$ is defined as:

$$f(\theta; b, \delta) = 2\delta \cdot Sigmoid\left(\frac{2(\theta - b)}{\delta}\right) + b - \delta \quad (28)$$

An example of the proposed stochastic variable transformation with $(b_1, \delta_1; b_2, \delta_2)_i = (0.5, 0.5; 1.5, 0.5)$ is shown in Figure 4.

As can be seen in Figure 4, the parts outside of $b_{i,1}$ and $b_{i,2}$ of the state space are compressed into the $\delta_{i,1}$ and $\delta_{i,2}$ widths of the parameter space, respectively, while the part between $b_{i,1}$ and $b_{i,2}$ is not affected.

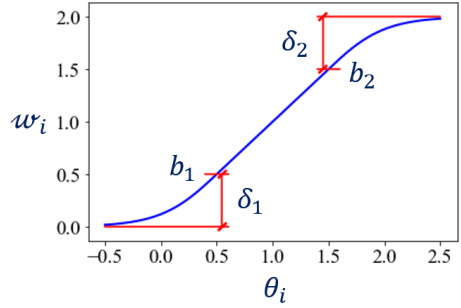

**Figure 4.** Example of stochastic variable transformation.

After applying this transformation to all $D$ parameters, the potential energy $U(\boldsymbol{\theta})$ for the new state $\boldsymbol{\theta}$ can be easily modified from the potential energy $U(\boldsymbol{w})$ of the parameter vector $\boldsymbol{w}$:

$$U(\boldsymbol{\theta}) = -\log p_{\boldsymbol{\theta}}(\boldsymbol{\theta}) + c = U(\boldsymbol{w}) - \sum_{i=1}^{D} T(\theta_i) + c' \qquad (29)$$

where $c'$ is an arbitrary constant to facilitate calculation, $-\sum_{i=1}^{D} T(\theta_i)$ is the modification term of the transformation, the potential energy of the parameter vector is given by Equation (2):

$$U(\boldsymbol{w}) = -\log(p(\mathcal{D}|\boldsymbol{w})) - \log(p(\boldsymbol{w})) + c^* \qquad (30)$$

where the elements of the parameter vector $\boldsymbol{w}$ can be derived from Equation (27).

As for the modification term, the function $T(\theta_i)$ is:

$$T(\theta_i) = \begin{cases} g(\theta_i; b_{i,1}, \delta_{i,1}), & \theta_i < b_{i,1} \\ 0, & b_{i,1} \leq \theta_i \leq b_{i,2} \\ g(\theta_i; b_{i,2}, \delta_{i,2}), & \theta_i > b_{i,2} \end{cases} \qquad (31)$$

where the function $g(\theta; b, \delta)$ is:

$$g(\theta; b, \delta) = 2 \log Sigmoid\left(\frac{2(\theta - b)}{\delta}\right) - \frac{2(\theta - b)}{\delta} + \log 4 \qquad (32)$$

The derivation is detailed in Appendix C.

Based on the new potential energy $U(\boldsymbol{\theta})$ calculated by Equations (27)-(32), the sampler using dynamics proposals can sample the whole real number space, and then transform into the bounded parameter space by Equation (27) to obtain the parameter samples exactly following the posterior PDF $p(\boldsymbol{w}|\boldsymbol{\mathcal{D}})$.

It can be seen that the proposed prior boundary treatment approach avoids the infinite gradient at the boundary and enables the application of bounded priors in MCMC methods with dynamics proposals, which is crucial for the free exploration in the early stage of network training in AM-SGHMC. There are several additional advantages of this approach: (1) The piecewise form of the transformation allows the boundaries on both ends to be treated separately. Furthermore, if there is no boundary on one end, the corresponding segment can be removed. (2) The identity transformation form of the middle segment ensures that the PDF shape of the state vector $\boldsymbol{\theta}$ at the high probability region remains the same as that of the parameter vector $\boldsymbol{w}$, which fits the needs of the scale-invariant input/output processing approach in section 3.1. (3) The involved nonlinear functions $Sigmoid(\cdot)$ and $\log Sigmoid(\cdot)$ are commonly used in neural networks, thus ensuring the stability and robustness of the computations.

This boundary treatment approach is specifically designed for component-wise prior boundaries, which are the most intuitive and commonly used. As for the application of this approach, if there is no prior boundary for some parameters that may interrupt the early stage of the training process, we can add them based on their parameter definitions or common sense, without affecting regions with reasonable parameter values. Then, for each prior boundary, the start point $b_i$ of corresponding transformation segment should be set inside and near the boundary, to keep the PDF shape at the high probability region unchanged. And the width $\delta_i$ is the distance between $b_i$ and the boundary.

# 4. AM-SGHMC ALGORITHM

According to the description in the previous section, the AM-SGHMC algorithm can be summarized as follows:

1. Initialize $z_0 = (\theta_0, p_0)$;

2. Repeat the following for $t = 0, \cdots, N - 1$:

In iteration $t$, let the most recent sample be $z_t = (\theta_t, p_t)$, then do the following to simulate a new sample $z_{t+1} = (\theta_{t+1}, p_{t+1})$:

a. Get sample information including $\theta_{t,i}$, $p_{t,i}$, $U(\theta_t)$, $\partial_{\theta_{t,i}} U(\theta_t)$, $Cate_i$ where $i = 1, \cdots, D$ ($U(\theta)$ is detailed in Section 3.3);

b. If in "burn-in" phase, update the estimation of the lower-order moments $\mu_U$ and $\sigma_U$ of potential energy $U(\theta)$ as well as standard deviation $\sigma_i, i = 1, \cdots, D$ of each dimension of the parameter posterior PDF (discussion of this adaptive estimation is presented in Appendix A);

c. Normalize the inputs of neural networks $\widehat{U}(\theta_t)$ and $\partial_{\theta_{t,i}} \widehat{U}^*(\theta_t)$ as in Equations (23)-(24);

d. Calculate elements of matrices $Q(z_t)$, $D(z_t)$ and $Q(\hat{z}_t)$ using the sampling strategy learned by neural networks to perform discretized dynamics simulation of Itô diffusion process defined by $Q(z)$ and $D(z)$ to obtain a new sample $z_{t+1} = (\theta_{t+1}, p_{t+1})$ as in Equations (12)-(19);

e. If training, recreate a copy of $z_{t+1}$ to stop the gradient flow;

f. If training, every $T_T$ steps, calculate the loss function defined in Equation (26) and update the neural network by back-propagation.

# 5. ILLUSTRATIVE EXAMPLES

## 5.1 Multi-story shear-building model

In order to verify the effectiveness and generalization ability of AM-SGHMC, we first consider some multi-story buildings excited by an earthquake. Three sets of noisy accelerometer data are simulated: Dataset 1 are the total acceleration data of a duration of 3 s collected from a 5-story building; Dataset 2 are the total acceleration data of a duration of 1 s collected from a 2-story building; Dataset 3 are the total acceleration data of a duration of 10 s collected from a 10-story building. The acceleration data are all collected with a sample interval of 0.01 s from the base, the first floor, and the roof and contaminated by a large amount of noise (i.e., 100% rms noise-to-signal ratio) as shown in Figure 5. In the data generating process, the ground motion is first generated using a zero-mean Gaussian distribution, and the structural parameters are randomly generated near their nominal values. Then, the structural response outputs are calculated and their average root-mean-square (rms) value is determined. Finally, the noise is generated using a zero-mean Gaussian distribution and added to the previously generated data, with the standard deviation being calculated as this rms value multiplied by the rms noise-to-signal ratio. Multi-story ($N$-story) linear shear building models are utilized for Bayesian inference and the stiffness parameter $k_i$ and damping coefficient $c_i$ for each story, $i = 1, \cdots, N$, of the models are estimated.

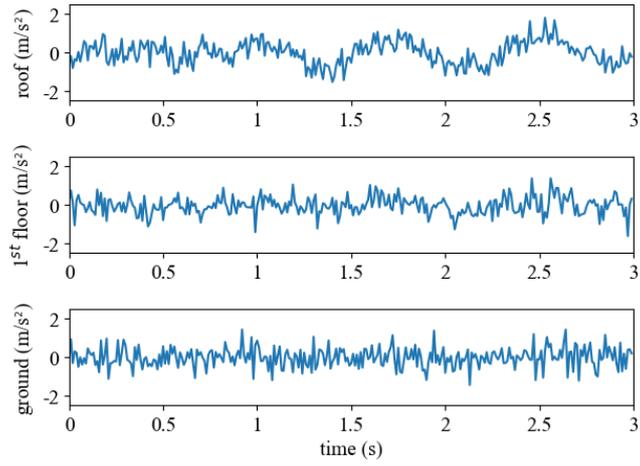

(a) Acceleration Dataset 1 (5-story).

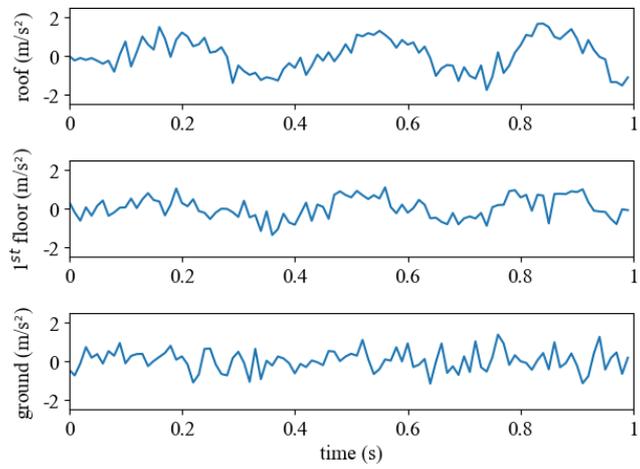

(b) Acceleration Dataset 2 (2-story).

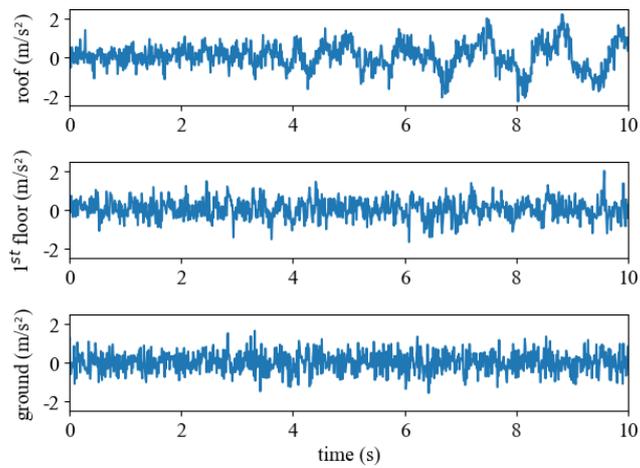

(c) Acceleration Dataset 3 (10-story).

**Figure 5.** Acceleration datasets in multi-story shear-building example.

The likelihood function $p(\mathcal{D}|\boldsymbol{w})$ is set the same as [8]. The number of observed degrees of freedom (2, i.e., first floor and roof) and the length of the discrete time history data (300, 100, 1000 for Datasets 1-3) are denoted as $N_o$ and $N_T$, respectively. For the $n$th observed degree of freedom at time $t_j$, the output predicted by the proposed structural model and the corresponding measured output are denoted as $y_n(t_j; \boldsymbol{w})$ and $\hat{y}_n(t_j)$, respectively. The prediction and measurement errors $\varepsilon_n(t_j) = \hat{y}_n(t_j) - y_n(t_j; \boldsymbol{w})$ for $n = 1, 2, \cdots, N_o$ and $j = 1, 2, \cdots, N_T$, are modeled as independent and identically distributed Gaussian variables with mean zero and some unknown variance $\sigma^2$, based on the Principle of Maximum Entropy [30-31]. Altogether, we need to estimate $D = 2N + 1$ model parameters with $\sigma$ included, and thus the likelihood function $p(\mathcal{D}|\boldsymbol{w})$ for this problem is:

$$p(\mathcal{D}|\boldsymbol{w}) = \frac{1}{(2\pi\sigma^2)^{\frac{N_o N_T}{2}}} \times \exp\left(-\frac{1}{2\sigma^2} \sum_{n=1}^{N_o} \sum_{j=1}^{N_T} [\hat{y}_n(t_j) - y_n(t_j; \boldsymbol{w})]^2\right) \quad (33)$$

Based on the nominal values $k_0 = 2 \times 10^7 \mathrm{Nm}^{-1}$, $c_0 = 6 \times 10^4 \mathrm{Nm}^{-1}\mathrm{s}$ (not equal to the exact values), and a roughly estimated error magnitude $\sigma_0 = 1.0 \mathrm{ms}^{-2}$, the uncertain parameters $w_i, i = 1, \cdots, 2N + 1$, are defined in dimensionless form: $w_i = k_i/k_0$ for $i = 1, \cdots, N$; $w_{N+i} = c_i/c_0$ for $i = 1, \cdots, N$; and $w_{2N+1} = \sigma/\sigma_0$. We set the prior PDF $p(\boldsymbol{w})$ as truncated independent distributions. The parameters $w_i = k_i/k_0$ and $w_{N+i} = c_i/c_0$ follow Gaussian distributions with means of 1 and coefficients of variation (c.o.v.) of 30%, truncated by intervals $[0.499, 1.501]$ and $[-0.502, 3.002]$, respectively, and $w_{2N+1} = \sigma/\sigma_0$ follows a lognormal distribution with median 1 and a logarithmic standard deviation of $s_0 = 0.3$ (the c.o.v. is about 30%), truncated by the interval $[0.098, 3.002]$.

Then the relationship between the parameter vector $\boldsymbol{w}$ and the state $\boldsymbol{\theta}$ of the

Markov chain is established according to Equations (27)-(28), with $(b_1, \delta_1; b_2, \delta_2)_{k_i} = (0.5, 0.001; 1.5, 0.001)$ for all parameters $w_i = k_i/k_0$, $(b_1, \delta_1; b_2, \delta_2)_{c_i} = (-0.5, 0.002; 3, 0.002)$ for all parameters $w_{N+i} = c_i/c_0$ and $(b_1, \delta_1; b_2, \delta_2)_\sigma = (0.1, 0.002; 3, 0.002)$ for parameter $w_{2N+1} = \sigma/\sigma_0$.

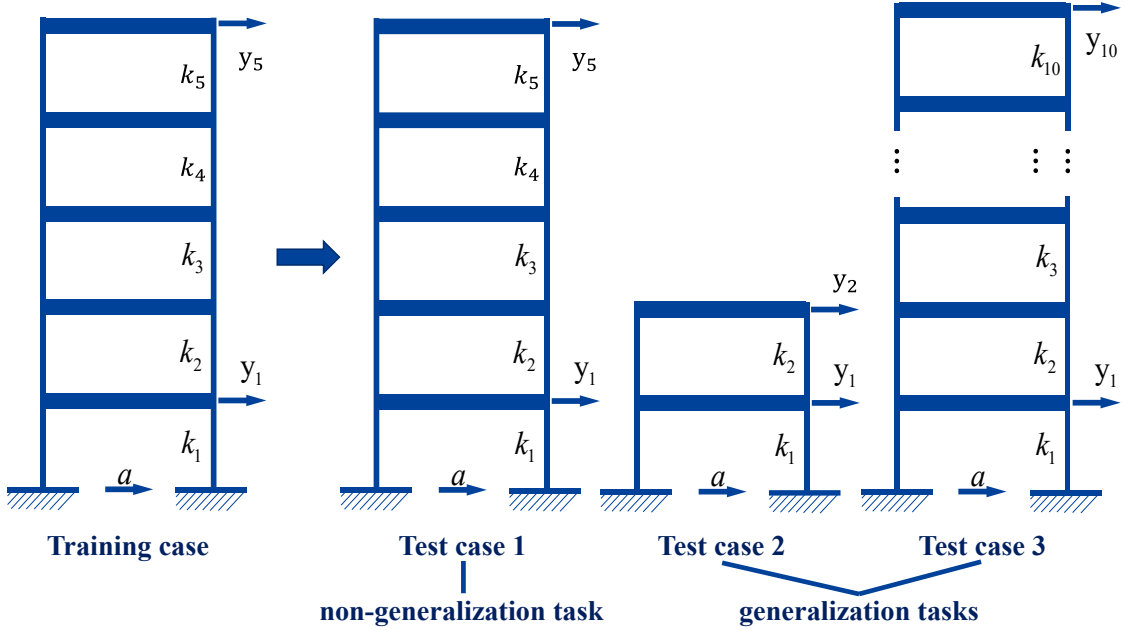

**Figure 6.** Experimental arrangement for multi-story shear-building example.

For AM-SGHMC, the architecture of two embedded neural networks $f_{\phi_Q}(\cdot)$ and $f_{\phi_D}(\cdot)$ and training setup are detailed in Appendix D. The experimental arrangement for this example is shown in Figure 6. The sampler is first trained by sampling on Dataset 1 (5-story), and then tested on Datasets 1-3 (5-story as non-generalization task, 2-story and 10-story as generalization tasks).

During each test of AM-SGHMC, $K = 32$ parallel chains for $T = 9000$ steps are simulated, where the first 3000 steps are burn in. The adaptive estimates are updated only during steps 300 to 2800 with $(\beta_1, \beta_2)_\theta = (0.99, 0.995)$ and $(\beta_1, \beta_2)_U = (0.99, 0.998)$ for parameter samples $\boldsymbol{\theta}$ and potential energy $U(\boldsymbol{\theta})$. The sampling trajectories (after burn in) for three cases by AM-SGHMC for some pair-wise parameters are shown in Figure 7 and compared with those by HMC method. For each test of HMC, similar with [8], the initial point is determined by 4000 steps of an efficient SPSA (simultaneous perturbation stochastic approximation) optimization algorithm, then $K = 32$ parallel chains for $T = 9000$ steps are simulated, where the first 500 steps are burn in. The colors in the figures are the two-dimensional marginal PDFs of the posterior PDFs estimated from the samples. The brighter the color, the higher the sample density is. As can be seen from the figure, the posterior sample distributions obtained by the two methods is similar, but the samples obtained by AM-SGHMC are more evenly distributed in the low probability part at the edge. As for the central high probability part obtained by HMC, the color of stiffness samples is less bright because many samples are wasted when exploring the low probability region, and the contour of damping samples is non-smooth because the initial exploration is relatively slow. The exact values of the dimensionless parameters used in the simulation are marked as pink stars in Figure 7. Comparing the sampling trajectories of the two methods, it can be seen that the deviation between the posterior sample distribution and the true value is mainly caused by the data, and has little to do with the difference of the two methods.

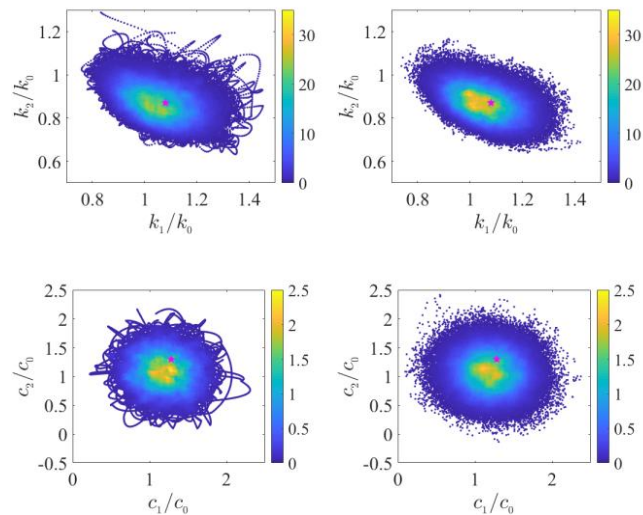

(a) Sampling trajectories of Dataset 1 (5-story).

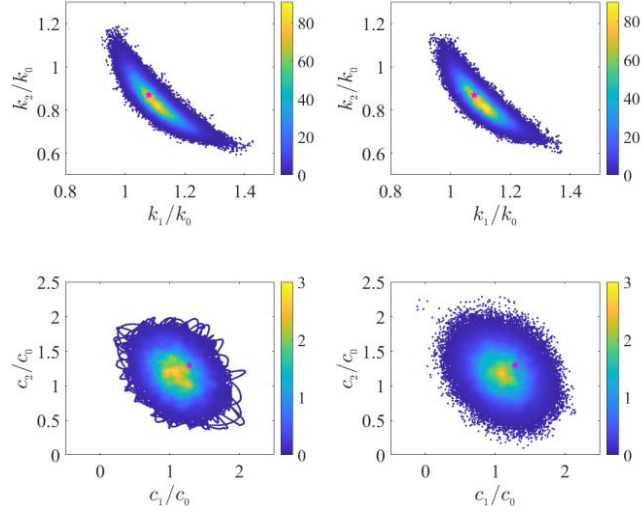

(b) Sampling trajectories of Dataset 2 (2-story).

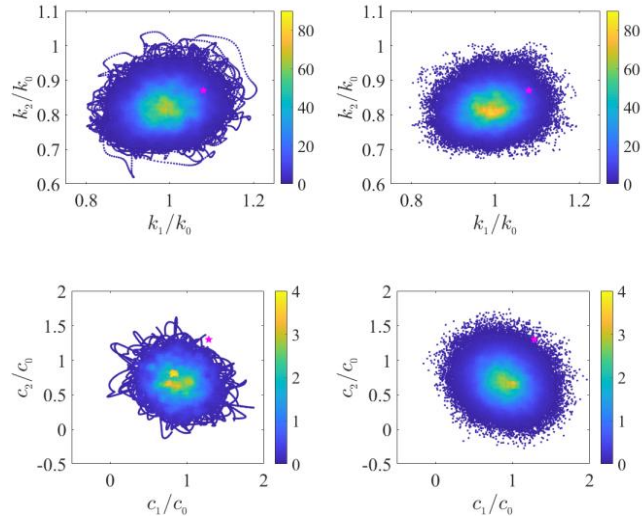

(c) Sampling trajectories of Dataset 3 (10-story).

**Figure 7.** Sampling trajectory plots for some pair-wise parameters by HMC (left column) and AM-SGHMC (right column) for three test cases.

It is worth clarifying that, due to the small number of measurement points, the posterior PDF is only strongly constrained by the data information in certain directions. This results in the shape of the two-dimensional projection resembling the prior distribution, an effect that is more pronounced in high-dimensional examples. Therefore, here we attempt to visualize the following two aspects: firstly, the significant difference in the scale of the posterior PDF across different directions; secondly, the nonlinear correlation between parameters within the high-dimensional space.

For the first aspect, which is also a main difficulty that classical HMC method attempts to solve [32-35], Principal Component Analysis (PCA) is used to find orthogonal and linearly independent directions of the samples, and to order the scales of the corresponding posterior PDF along these directions. Figure 8 shows the two-dimensional projections of the samples obtained by AM-SGHMC algorithm in the directions of the first and last principal components (PCs) of the three cases, and the projections of the samples obtained by the HMC algorithm in the same directions are also shown for comparison. To facilitate comparison, the sample intervals of the AM-SGHMC algorithm are accordingly increased so that the sample numbers of the two methods in the three cases are similar.

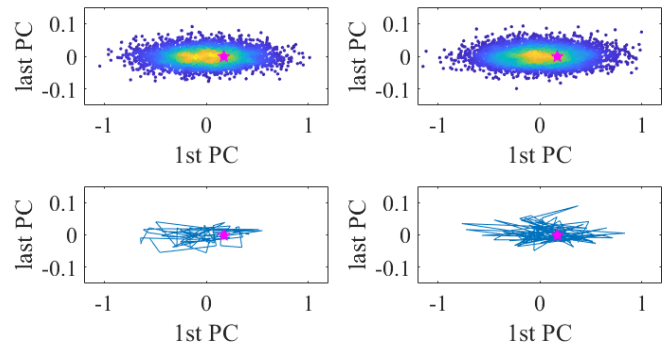

(a) Projection from test case 1 (5-story).

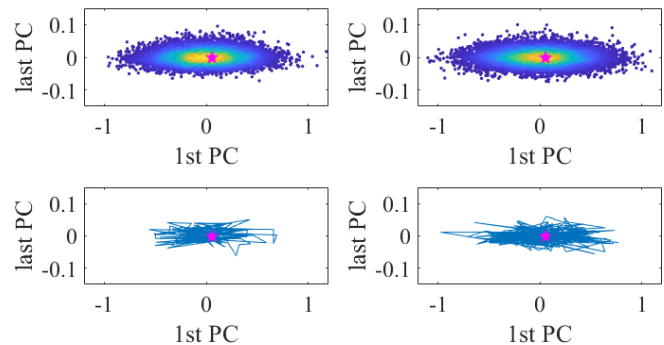

(b) Projection from test case 2 (2-story).

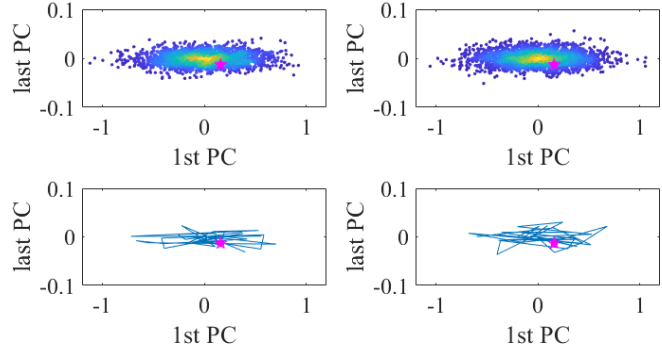

(c) Projection from test case 3 (10-story).

**Figure 8.** Projection of samples obtained by HMC (left column) and AM-SGHMC (right column) in the first and last PC directions for three test cases.

The figures at the top of each case show the samples from all 32 chains. The colors in these figures correspond to the two-dimensional marginal PDF of the posterior PDF estimated from the samples. It can be seen that the overall distribution obtained by the two methods is consistent, and even though the projection will obscure the high-dimensional shape and make the vertical scale larger, the horizontal and vertical scales are still an order of magnitude different. Each figure at the bottom of each case shows samples from only one chain. It can be seen that in test cases 1 and 2, AM-SGHMC algorithm is more effective for the exploration in the direction of the first PC; in test case 3, due to the large sampling interval required by HMC, the exploration efficiency of the two methods is similar.

To illustrate how projection transformation obscures the geometric characteristic of high-dimensional samples, consider the ideal distribution of their two-dimensional projection in the figure:

$$p\left(\theta_{P_i}, \theta_{P_j}|\mathcal{D}\right) = \int p(\boldsymbol{\theta}|\mathcal{D}) \, d\boldsymbol{\theta}_{\overline{P_{i,j}}} = \int p\left(\theta_{P_i}, \theta_{P_j}|\boldsymbol{\theta}_{\overline{P_{i,j}}}; \mathcal{D}\right) p\left(\boldsymbol{\theta}_{\overline{P_{i,j}}}|\mathcal{D}\right) d\boldsymbol{\theta}_{\overline{P_{i,j}}} \quad (34)$$

where $\theta_{P_i}$, $\theta_{P_j}$, and $\boldsymbol{\theta}_{\overline{P_{i,j}}}$ are the components of the parameter in the $i$-th, $j$-th and all other PC directions, $P_i$, $P_j$ and $\overline{P_{i,j}}$, respectively. When there is a high-dimensional correlation, the randomness of the unknown correlation components causes the conditional PDF $p\left(\theta_{P_i}, \theta_{P_j} | \boldsymbol{\theta}_{\overline{P_i P_j}}; \mathcal{D}\right)$ to change, making the two-dimensional projection PDF become the weighted average of a series of continuously varying conditional PDFs, and finally form a relatively smooth distribution in the figure. Since the PCs extracted by PCA are linearly independent, when the conditional PDF varies enough to obscure its own shape, the two-dimensional projection in the figure appears to be close to the Gaussian distribution, which is more common in higher dimensional problems.

To visualize the high-dimensional nonlinear correlations between parameters in the posterior PDFs, the following three-dimensional relationships are shown. Under the conditions of the selected two-dimensional PC values $\theta_{P_i}$ and $\theta_{P_j}$, consider the posterior mean of a third dimensional PC $\theta_{P_k}$ of the parameters, which can be approximated using Monte Carlo method with $N_\delta$ samples $\theta_{n,P_k}, n = 1, \cdots, N_\delta$:

$$\mathbb{E}\left(\theta_{P_k} | \theta_{P_i}, \theta_{P_j}\right) = \int \theta_{P_k} p\left(\theta_{P_k} | \theta_{P_i}, \theta_{P_j}\right) d\theta_{P_k} \approx \frac{1}{N_\delta} \sum_{\substack{n=1 \\ \theta_{n,P_k} \\ \sim p\left(\theta_{P_k} | \theta_{P_i}, \theta_{P_j}\right)}}^{N_\delta} \theta_{n,P_k} \quad (35)$$

However, based on the sampling results presented earlier, due to the absence of multiple $\theta_{n,P_k}$ samples with exactly the same $\left(\theta_{P_i}, \theta_{P_j}\right)$ values, a further approximation is made by utilizing $\theta_{n,P_k}$ samples within the neighborhood of $\left(\theta_{P_i}, \theta_{P_j}\right)$ and applying multiple iterations of averaging to mitigate noise. Specifically, samples within a threshold of 0.3 standard deviations are used, and 3 iterations of averaging are performed. This iterative operation can retain the overall trend of the conditional posterior mean $\mathbb{E}\left(\theta_{P_k} | \theta_{P_i}, \theta_{P_j}\right)$ changing with conditions $\theta_{P_i}$ and $\theta_{P_j}$.

Two obvious trends from test case 3 are selected and shown in Figure 9. The colors in the figure represent the two-dimensional marginal PDF of the posterior PDF, estimated from the samples, along the $P_i$ and $P_j$ directions.

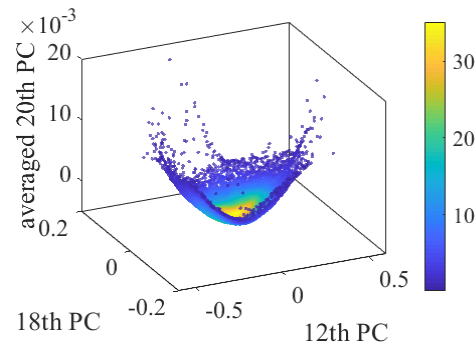

(a) Nonlinear relationship close to the paraboloid.

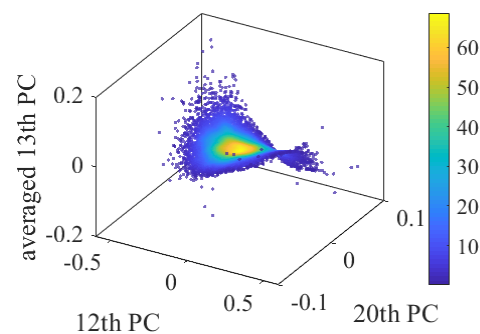

(b) Nonlinear relationship close to the saddle surface.

**Figure 9.** Top (left column) and oblique (right column) view of two obvious nonlinear relations obtained by filtering in test case 3.

As can be seen from the figures, even if the high-dimensional samples are compressed into three-dimensional space by projection, the nonlinear relationship between the parameters can still be seen after being smoothed by iterative averaging. It is the high-dimensional complex manifold behind this nonlinear phenomenon that leads to the main difficulties faced by the sampling methods.

Back to the sampling results, to further confirm the consistency of the results of the two methods, we use the naive version loss function in Equation (20). As mentioned in Section 3.2, this naive version sample-based loss is independent of the sample order, so that it directly indicates the correctness of the local distribution of samples. The Naive Loss of the two methods on Datasets 1-3 are evaluated and recorded in Tables 1-3, respectively. It can be seen that the mean Naive Loss values of the two methods are always similar, which indicates the consistency of sample distributions obtained by the two methods.

In order to compare the performance of the different samplers, Effective Sample Size (ESS) is employed, which is often used in the evaluation of samplers. ESS for a sequence of correlated samples can be viewed as the number of independent samples generated by the same PDF, where the two sets of samples can achieve the same accuracy in estimating the mean of the PDF. Our implementation of ESS estimation follows [23] and is briefly presented here.

For a set of correlated samples $\mathbf{\Theta}_T = \{\boldsymbol{\theta}_\tau\}_{\tau=1}^T$ generated by a Markov chain, its ESS is estimated by:

$$\text{ESS}(\mathbf{\Theta}_T) = \frac{T}{1 + 2\sum_{s=1}^{\left[\frac{T}{3}\right]-1}\left(1 - \frac{s}{T}\right)\frac{\rho_s}{\rho_0}} \tag{36}$$

where $\frac{\rho_s}{\rho_0}$ is the autocorrelation of $\mathbf{\Theta}_T$ at lag $s$ estimated by:

$$\rho_s = \frac{1}{T-s}\sum_{\tau=s+1}^{T}(\boldsymbol{\theta}_\tau - \widehat{\boldsymbol{\mu}})^\text{T}(\boldsymbol{\theta}_{\tau-s} - \widehat{\boldsymbol{\mu}}) \tag{37}$$

where $\widehat{\boldsymbol{\mu}}$ is estimated by:

$$\widehat{\boldsymbol{\mu}} = \frac{1}{T}\sum_{\tau=1}^{T}\boldsymbol{\theta}_\tau \tag{38}$$

The sum in Equation (36) is truncated whenever $s = 1000$ or $s$ is even number and $\rho_{s-1} + \rho_s < 0$.

The ESSs for both methods on each Dataset are evaluated. With each corresponding time consumption recorded, we can define the sampling efficiency of the samplers as ESS per hour (ESS/h). The sampling efficiency of the two methods on Datasets 1-3 are also tabulated in Tables 1-3, respectively.

As can be seen from the tables, the efficiency of AM-SGHMC is 3.2, 2.4 and 4.6 times that of HMC in 5-story, 2-story and 10-story tasks, respectively, and the higher the parameter dimension, the more the improvement is. These results show that AM-SGHMC has the generalization ability and is more advantageous in higher-dimensional sampling problems.

**Table 1.** Sampling performance comparison of Dataset 1 (5-story).

| Methods | Naive Loss | ESS | Time (h) | ESS/h |
| --- | --- | --- | --- | --- |
| HMC | 296.01 | 64.53 | 0.35 | 184.37 |
| AM-SGHMC | 294.93 | 198.79 | 0.33 | **602.39** |

**Table 2.** Sampling performance comparison of Dataset 2 (2-story).

| Methods | Naive Loss | ESS | Time (h) | ESS/h |
| --- | --- | --- | --- | --- |
| HMC | 83.06 | 158.67 | 0.20 | 793.35 |
| AM-SGHMC | 82.23 | 425.18 | 0.22 | **1932.6** |

**Table 3.** Sampling performance comparison of Dataset 3 (10-story).

| Methods | Naive Loss | ESS | Time (h) | ESS/h |
| --- | --- | --- | --- | --- |
| HMC | 1021.68 | 37.84 | 0.88 | 43.00 |
| AM-SGHMC | 1022.92 | 146.82 | 0.73 | **201.12** |

## *5.2 Multi-story braced-frame building model*

In this section, a more detailed building model is employed to further verify the ability of AM-SGHMC. We still consider the $N$-story buildings excited by an earthquake. Different from the previous example, which considers only one degree of freedom (DOF) per floor, in this example, with reference to the benchmark structure [36-

[39] proposed by IASC (International Association for Structural Control) -ASCE (American Society of Civil Engineers) Task Group on SHM, we consider the planar motion of 3 DOF for each floor, so that a $3N$-DOF model is obtained by assuming that the floor slabs of the braced frame "are rigid in and out of the plane". The 4-story 2-bay by 2-bay steel braced-frame IASC-ASCE benchmark test structure is shown in Figure 10. The structural layout of the members on each floor of the benchmark structure is the same, and is adopted by the $N$-story buildings in this example.

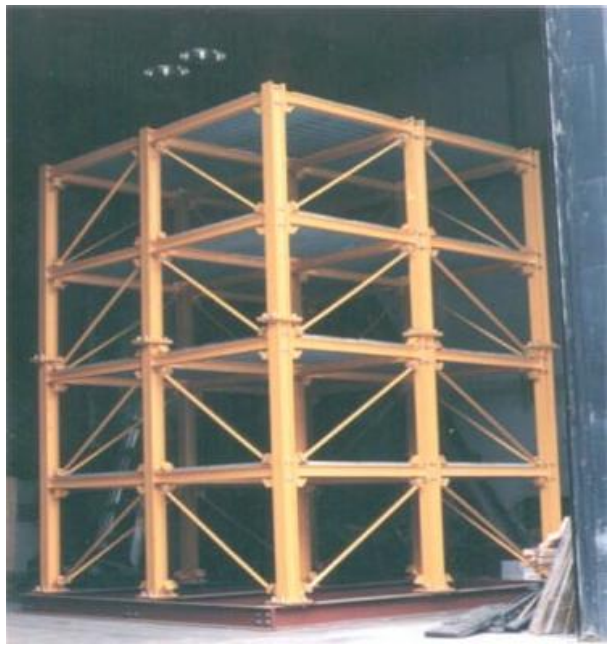

**Figure 10.** The 4-story IASC-ASCE benchmark test structure [36].

Four sets of noisy accelerometer data are simulated: Dataset 1 are training data collected from a 4-story building contaminated by a typical amount of noise (i.e., 10% rms noise-to-signal ratio); Datasets 2-4 are test data collected from a 2-story building, a 4-story building, and a 6-story building, respectively, contaminated by a large amount of noise (i.e., 50% rms noise-to-signal ratio). The total acceleration data are all collected from the base, the first floor, and the roof along the horizontal direction of each outer wall, with a sample interval of 0.01s and a duration of 3 s, as shown in Figure 11.

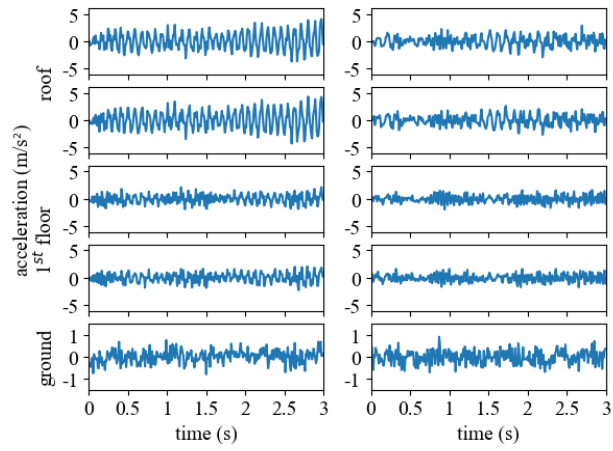

(a) Acceleration Dataset 1 (4-story with typical noise).

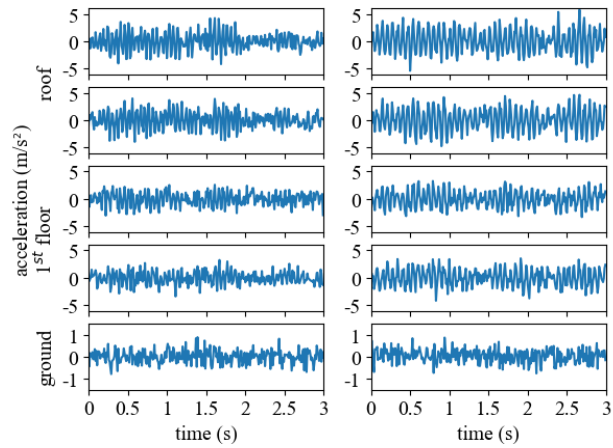

(b) Acceleration Dataset 2 (2-story with large noise).

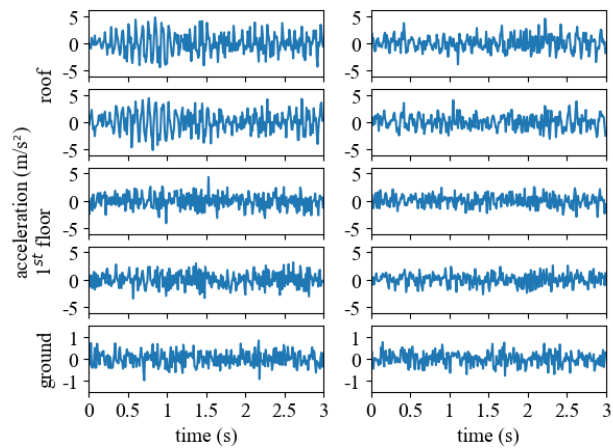

(c) Acceleration Dataset 3 (4-story with large noise).

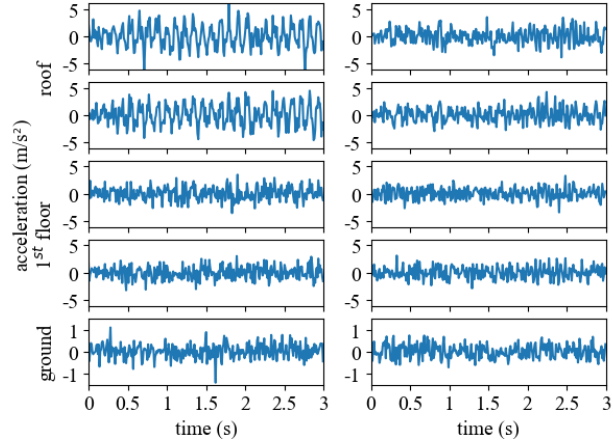

(d) Acceleration Dataset 4 (6-story with large noise).

**Figure 11.** Acceleration datasets in multi-story braced-frame building example

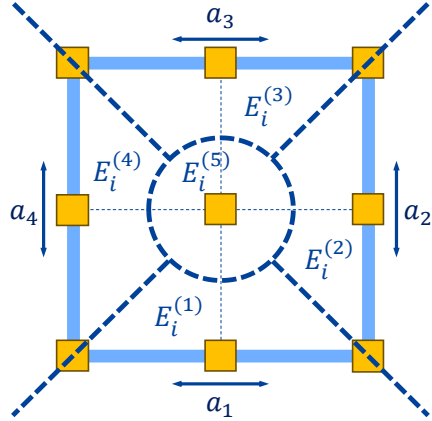

**Figure 12.** Each floor is divided into 5 regions by dotted lines

The $3N$-DOF braced-frame building models are utilized for Bayesian inference and the Young's modulus parameters $E_i^{(j)}$ of the 5 regions, $j = 1, \cdots, 5$, as shown in Figure 12, for each story, $i = 1, \cdots, N$, of the models are estimated. Specifically, each component takes the Young's modulus of the corresponding region, while the corner columns at the junction of two regions will use the average of the Young's modulus of the two regions. The likelihood function $p(\mathcal{D}|w)$ is set the same as the previous example, as shown in Equation (33), but with $N_o = 8$, $N_T = 300$ and the $3N$-DOF braced-frame building model.

Based on the nominal value $E_0 = 2 \times 10^{11}$ Pa (not equal to the exact value), and a roughly estimated error magnitude $\sigma_0 = 0.8 \text{ms}^{-2}$, the uncertain parameters $w_k, k = 1, \cdots, 5N + 1$, are defined in dimensionless form: $w_{i \times 5+j-5} = E_i^{(j)}/E_0$ for $i = 1, \cdots, N; j = 1, \cdots, 5$; and $w_{5N+1} = \sigma/\sigma_0$. We set the prior PDF $p(\boldsymbol{w})$ as truncated independent distributions. The dimensionless modulus of each component determined by parameters $w_k = E_i^{(j)}/E_0, k = 1, \cdots, 5N$ follow Gaussian distributions with means of 1 and coefficients of variation (c.o.v.) of 10%, truncated by interval $[0.1, 2.0]$, and $w_{5N+1} = \sigma/\sigma_0$ follows a lognormal distribution with median 1 and a logarithmic standard deviation of $s_0 = 0.3$ (the c.o.v. is about 30%), also truncated by interval $[0.1, 2.0]$. Then the relationship between the parameter vector $\boldsymbol{w}$ and the state $\boldsymbol{\theta}$ of the Markov chain is established according to Equations (27)-(28), with $(b_1, \delta_1; b_2, \delta_2) = (0.2, 0.1; 1.9, 0.1)$ for all parameters $w_i$.

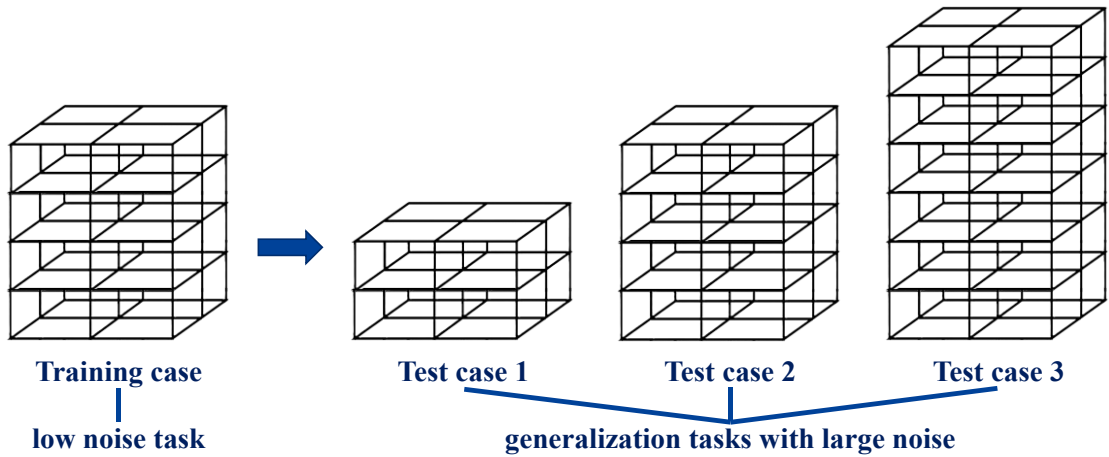

**Figure 13.** Experimental arrangement for multi-story braced-frame building example.

For AM-SGHMC, the architecture of two embedded neural networks $f_{\phi_Q}(\cdot)$ and $f_{\phi_D}(\cdot)$ and training setup are also detailed in Appendix D. The experimental arrangement for this example is shown in Figure 13. The sampler is firstly trained by sampling on Dataset 1 (4-story with typical noise), and then tested on Datasets 2-4 (2-

story, 4-story and 6-story with large noise as generalization tasks).

During each test of AM-SGHMC, $K = 32$ parallel chains for $T = 9000$ steps are simulated, where the first 3000 steps are burn in. The adaptive estimates are updated only during steps 500 to 2800 with $(\beta_1, \beta_2)_\theta = (0.99, 0.998)$ and $(\beta_1, \beta_2)_U = (0.99, 0.998)$ for parameter samples $\boldsymbol{\theta}$ and potential energy $U(\boldsymbol{\theta})$. The samples (after burn in) for three cases by AM-SGHMC for some pair-wise parameters are shown in Figure 14 and compared with those by HMC method. The colors in the figures are the two-dimensional marginal PDFs of the posterior PDFs estimated from the samples. As can be seen from the figure, similar to the previous example, the posterior sample distributions obtained by the two methods is similar, but the samples obtained by AM-SGHMC are more evenly distributed, resulting in smoother edges between different colors. As mentioned in the previous section, for such higher-dimensional problems, the shapes of the two-dimensional projections are very similar to those of the prior distributions.

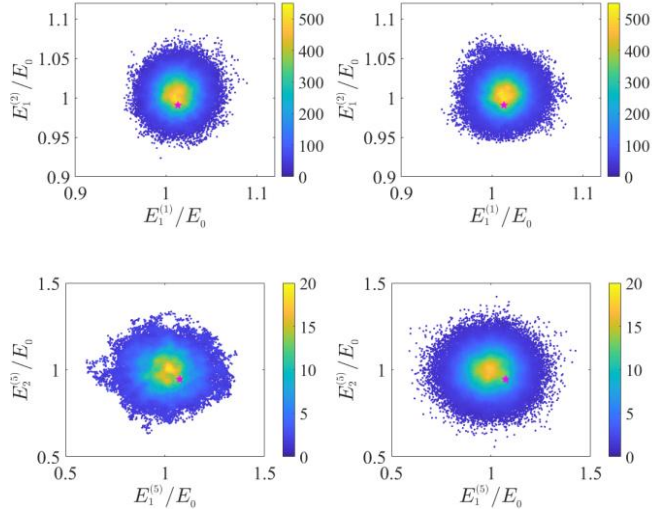

(a) Samples of Dataset 2 (2-story).

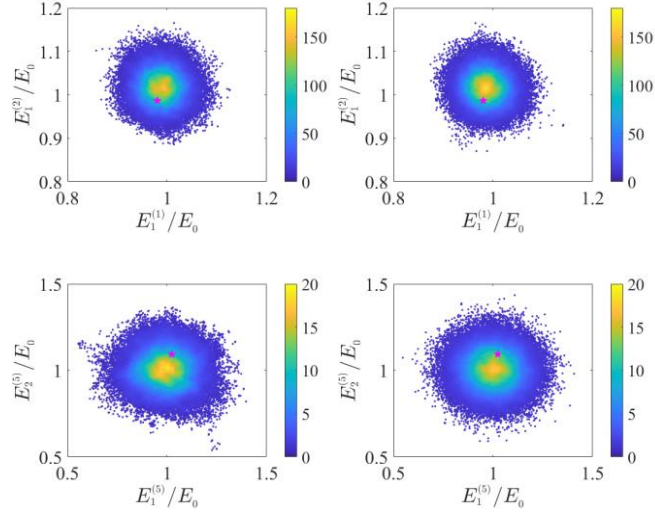

(b) Samples of Dataset 3 (4-story).

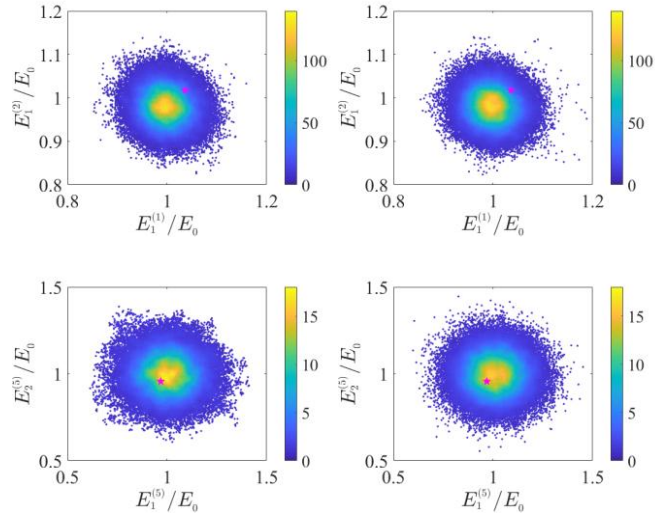

(c) Samples of Dataset 4 (6-story).

**Figure 14.** Sample plots for some pair-wise parameters by HMC (left column) and AM-SGHMC (right column) for three test cases.

Similar to the previous example, to prove and visualize the high-dimensional nonlinear correlations between parameters in the posterior PDFs, two obvious nonlinear relations obtained by conditional averaging from test case 2 are selected and shown in Figure 15. The colors in the figure represent the two-dimensional marginal PDF of the posterior PDF, estimated from the samples, along the two conditional PC directions.

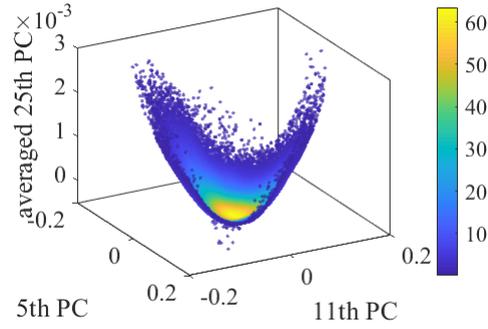

(a) Nonlinear relationship close to the paraboloid.

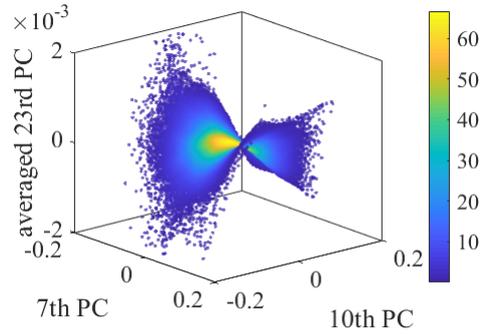

(b) Nonlinear relationship close to the saddle surface.

**Figure 15.** Oblique view of two obvious nonlinear relations obtained by filtering in test case 2.

As in the previous example, to compare the performance of the different samplers, the Naive Loss, Effective Sample Size (ESS) and sampling efficiency (ESS/h) for both methods on each of the datasets are evaluated and tabulated in Tables 4-6, respectively.

**Table 4.** Sampling performance comparison of Dataset 2 (2-story).

| Methods | Naive Loss | ESS | Time (h) | ESS/h |
|---|---|---|---|---|
| HMC | 2635.97 | 7.96 | 3.20 | 2.49 |
| AM-SGHMC | 2636.83 | 31.32 | 1.89 | **16.59** |

**Table 5.** Sampling performance comparison of Dataset 3 (4-story).

| Methods | Naive Loss | ESS | Time (h) | ESS/h |
|---|---|---|---|---|
| HMC | 2835.20 | 13.37 | 5.69 | 2.35 |
| AM-SGHMC | 2835.10 | 60.33 | 3.18 | **18.99** |

**Table 6.** Sampling performance comparison of Dataset 4 (6-story).

| Methods | Naive Loss | ESS | Time (h) | ESS/h |
|---|---|---|---|---|
| HMC | 2898.67 | 14.24 | 8.29 | 1.71 |
| AM-SGHMC | 2898.09 | 69.89 | 4.59 | **15.23** |

As can be seen from the tables, the similar Naive Loss values indicate the similar sample distributions obtained by the two methods. As for the sampling efficiency, similar to the previous example, the sampling efficiency of AM-SGHMC is 6.66, 8.08 and 8.87 times that of HMC in 2-story, 4-story and 6-story tasks, respectively, and the higher the parameter dimension, the more the improvement is. These results show again the generalization ability of AM-SGHMC and its advantages in higher-dimensional sampling problems.

## 6. CONCLUDING REMARKS

In recent decades, MCMC methods have been extensively used for Bayesian updating of structural dynamic models in structural health monitoring. Recently, neural network-enhanced MCMC algorithms were developed to improve performance for specific Bayesian updating tasks. However, a key challenge is the need for retraining the embedded neural networks for new tasks, which is time-consuming and thereby diminishes their competitiveness.

In this paper, a novel meta-learning stochastic simulation approach, called the AM-SGHMC, is developed to implement meta-learning for Bayesian updating of structural dynamic models where the MCMC sampler is based on trained adaptive neural networks that utilize the characteristics of the various posterior PDFs that are involved. Specifically, an input/output processing approach is proposed that takes the parameter types into consideration and guarantees scale invariance with respect to the posterior PDF of the model parameters. This approach can improve the generalization ability of the sampler because the shape of the conditional posterior PDF for parameters of the same type will be relatively similar than those of different types. In addition, the loss function and its back-propagation path are improved so that they are more suitable for the training of networks in Markov chain environments.

An approach is also proposed to enable the application of bounded priors in MCMC methods with dynamics proposals. Bounded priors, such as uniform distributions and truncated Gaussian distributions, can limit the parameters to a reasonable range, so as to avoid complex low probability regions. However, the infinite gradient at the boundary cannot be sensed by dynamics proposals. The proposed approach avoids this problem by performing carefully designed stochastic variable transformations that extend the parameter space near the boundary to infinity, which is crucial for free exploration in the early stage of network training.

Two examples involving Bayesian updating of a multi-story shear-building model and a multi-story braced-frame building model under ground motions are used to demonstrate the effectiveness and generalization ability of AM-SGHMC. Specifically, in the shear-building model example, the adaptive meta-learning sampler of AM-SGHMC is trained on a five-story building model updating task and then tested on a five-story (non-generalization), a two-story (generalization) and a ten-story (generalization) tasks. The test results show that, with similar accuracy, the sampling efficiency (Effective Sample Size per hour, ESS/h) of AM-SGHMC is 3.2 times, 2.4 times and 4.6 times, respectively, that of the classical Hamiltonian Monte Carlo method (HMC) for Bayesian updating of structural dynamic models. In the braced-frame building model example, the AM-SGHMC sampler is trained on a four-story building model updating task with typical amounts of noise, and then tested on a two-story, a four-story and a six-story generalization tasks with larger noise levels. The test results show that, with similar accuracy, the sampling efficiency (ESS/h) of AM-SGHMC is 6.6 times, 8.0 times and 8.8 times, respectively, of that of the HMC method. In these examples, due to savings in training time, the computational effort for AM-SGHMC is less than that of the HMC method. Note that even though both examples involve linear structural dynamics, their relationships among the parameters indicated by the posterior PDFs are nonlinear, which can be seen intuitively from Figs. 9 and 15, indicating their essential system identification problems are actually also nonlinear. Therefore, there is no fundamental theoretical barrier for the application to nonlinear structural system identification problems, as long as there is a suitable model class that includes a likelihood function and a prior PDF of its parameters.

**Declaration of competing interest**

The authors declare that they have no known competing financial interests or personal relationships that could have appeared to influence the work reported in this paper.

**Acknowledgements**

The study was supported by National Natural Science Foundation of China under Grant No. 52078174.

# Appendix A  *Adaptive estimate process for AM-SGHMC*

The adaptive estimate process mentioned in the main text is an extension of the lower-order moment estimate process adopted in Adam [29], one of the most popular optimizers.

Adam implements optimization based on adaptive estimates of the 1st moment and raw 2nd moment of the first-order gradient, where the gradients $g_t, t = 1, \cdots, T$ can only be obtained at the corresponding timestep $t$ based on the parameters updated in the previous step. The same situation occurs when we estimate the first two moments of the parameters $\boldsymbol{\theta}$ and potential energy $U(\boldsymbol{\theta})$ during sampling, that is, $\boldsymbol{\theta}_t$ and $U(\boldsymbol{\theta}_t)$ are obtained at successive timesteps $t = 1, \cdots, T$. Therefore, in this section, the samples, such as $g_t$, $\boldsymbol{\theta}_t$, and $U(\boldsymbol{\theta}_t)$, obtained for estimating the first two moments are uniformly denoted as $y_t$ to focus on the adaptive estimate process.

The lower-order moment estimate process adopted in Adam is extracted and given in Algorithm 1. In this algorithm, two exponential moving averages are used to update biased 1st and raw 2nd moments, $m_t$ and $v_t$, with exponential decay rates $\beta_1$ and $\beta_2$ (recommended setting $\beta_1 = 0.9$, $\beta_2 = 0.999$), respectively. Initialization bias correction terms are subsequently utilized to correct the bias caused by initializing the running average with zeros.

---

**Algorithm 1:** First-two-moment estimate in Adam. All operations on vectors are elementwise.

---

**Require:** $\beta_1, \beta_2 \in [0,1)$: Exponential decay rates for the moment estimates
**Require:** $y_1, \cdots, y_T$: Samples obtained at subsequent timesteps $t = 1, \cdots, T$

$\quad m_0 \leftarrow 0$ (Initialize 1st moment vector)
$\quad v_0 \leftarrow 0$ (Initialize 2nd moment vector)
$\quad t \leftarrow 0$ (Initialize timestep)
$\quad$ **for** $t' = 1, \cdots, T$ **do**
$\quad\quad t \leftarrow t + 1$
$\quad\quad m_t \leftarrow \beta_1 m_{t-1} + (1 - \beta_1) y_t$ (Update biased 1st moment estimate)
$\quad\quad v_t \leftarrow \beta_2 v_{t-1} + (1 - \beta_2) y_t^2$ (Update biased raw 2nd moment estimate)
$\quad\quad \widehat{m}_t \leftarrow m_t / (1 - \beta_1^t)$ (Compute bias-corrected 1st moment estimate)

---

$\hat{v}_t \leftarrow v_t/(1-\beta_2^t)$ (Compute bias-corrected raw 2nd moment estimate)
**end for**
**return** $\hat{m}_t, \hat{v}_t$ at timesteps $t = 1, \cdots, T$ (Estimating the first two moments)

The application scenario of the estimation processes in our method differs from that in Adam in two main aspects. On the one hand, when sampling by MCMC method, $K$ parallel Markov chains are simulated at the same time, which means that at each timestep $t$, a batch of samples $\{y_{t,k}\}_{k=1}^{K}$ of size $K$ will be obtained instead of one sample. On the other hand, because the means of the parameters $\boldsymbol{\theta}$ and potential energy $U(\boldsymbol{\theta})$ are not zero, the centered second moment is a more accurate estimate of the variance than the raw second moment. Considering these differences, the iterative update formulas in Algorithm 1 need to be modified accordingly.

Same as Algorithm 1, all operations on vectors in this section are elementwise. We start with the basic statistical formulas for the mean and the variance. When the batch size is $K$, the mean $\mu_t$ and the variance $\sigma_t^2$ at timestep $t$ can be calculated as:

$$\mu_t = \frac{1}{tK} \sum_{i=1}^{t} \sum_{k=1}^{K} y_{i,k} \tag{39}$$

$$\sigma_t^2 = \frac{1}{tK-1} \sum_{i=1}^{t} \sum_{k=1}^{K} (y_{i,k} - \mu_t)^2 \tag{40}$$

According to these general term formulas, the recurrence formulas of these two series of numbers can be obtained:

$$\mu_t = \frac{t-1}{t} \mu_{t-1} + \frac{1}{t}\left[\frac{1}{K}\sum_{k=1}^{K} y_{t,k}\right] \tag{41}$$

$$\sigma_t^2 = \frac{tK-K-1}{tK-1} \sigma_{t-1}^2 + \frac{K}{tK-1}\left[\frac{1}{K}\sum_{k=1}^{K}(y_{t,k}-\mu_t)^2\right] + \frac{tK-K}{tK-1}(\mu_t - \mu_{t-1})^2 \tag{42}$$

Then by defining $\beta_1$ and $\beta_2$ as:

$$\beta_1 = \frac{t-1}{t} \tag{43}$$

$$\beta_2 = \frac{tK - K - 1}{tK - 1} \tag{44}$$

Equations (41)-(42) can be finally rewritten as:

$$\mu_t = \beta_1 \mu_{t-1} + (1 - \beta_1)\left[\frac{1}{K}\sum_{k=1}^{K} y_{t,k}\right] \tag{45}$$

$$\sigma_t^2 = \beta_2 \sigma_{t-1}^2 + (1 - \beta_2)\left[\frac{1}{K}\sum_{k=1}^{K}(y_{t,k} - \mu_t)^2\right]$$

$$+ \left[\beta_2 + (1-\beta_2)\frac{1}{K}\right](\mu_t - \mu_{t-1})^2 \tag{46}$$

Based on the recurrence formulas (45)-(46), we modified Algorithm 1 into our extended version as shown in Algorithm 2.

---

**Algorithm 2:** First-two-moment estimate with batch input and non-zero mean. All operations on vectors are elementwise.

---

**Require:** $\beta_1, \beta_2 \in [0,1)$: Exponential decay rates for the moment estimates
**Require:** $K$: Batch size of samples obtained at each timestep
**Require:** $\{y_{1,k}\}_{k=1}^{K}, \cdots, \{y_{T,k}\}_{k=1}^{K}$: Samples obtained at subsequent timesteps $t = 1, \cdots, T$

$m_0 \leftarrow 0$ (Initialize 1st moment vector)
$v_0 \leftarrow 0$ (Initialize 2nd moment vector)
$t \leftarrow 0$ (Initialize timestep)
**for** $t' = 1, \cdots, T$ **do**
  $t \leftarrow t + 1$
  $\bar{y}_t \leftarrow \frac{1}{K}\sum_{k=1}^{K} y_{t,k}$
  $m_t \leftarrow \beta_1 m_{t-1} + (1 - \beta_1)\bar{y}_t$ (Update biased 1st moment estimate)
  $\hat{m}_t \leftarrow m_t/(1 - \beta_1^t)$ (Compute bias-corrected 1st moment estimate)
  **if** $t = 1$ **do**
    $\hat{m}_0 \leftarrow \hat{m}_1$
  **end if**
  $v'_{t-1} \leftarrow (\hat{m}_t - \hat{m}_{t-1})^2 + v_{t-1}$
  $\overline{y_t^2} \leftarrow \frac{1}{K}\left[(\hat{m}_t - \hat{m}_{t-1})^2 + \sum_{k=1}^{K}(y_{t,k} - \hat{m}_t)^2\right]$

$v_t \leftarrow \beta_2 v'_{t-1} + (1-\beta_2)\overline{y_t^2}$ (Update biased centered 2nd moment estimate)
$\hat{v}_t \leftarrow v_t/(1-\beta_2^t)$ (Compute bias-corrected centered 2nd moment estimate)
**end for**
**return** $\hat{m}_t, \hat{v}_t$ at timesteps $t = 1, \cdots, T$ (Estimating the first two moments)

---

In AM-SGHMC, the proposed first-two-moment estimate process shown in Algorithm 2 is used in the adaptive estimation of the lower-order moments $\mu_U$ and $\sigma_U^2$ of the potential energy $U(\boldsymbol{\theta})$.

As for the adaptive estimation of the variances $\sigma_i^2, i = 1, \cdots, D$ of each component of the parameter samples during the middle part of the burn-in, the situation is somewhat different.

In the testing process, based on the estimation in the training process, an initial rough estimate of the variances can usually be made based on the qualitative and quantitative changes in the new data, which can be used as a prior initial value vector of the adaptive estimation.

Denoting the prior initial value vector as $v_0^*$, and set $v_0 \leftarrow v_0^*$, the bias correction term $\hat{v}_t \leftarrow v_t/(1-\beta_2^t)$ in Algorithm 2 should be modified as:

$$\hat{v}_t = \frac{v_t - \beta_2^t v_0^*}{1 - \beta_2^t} \tag{47}$$

However, same as the initialization bias correction term utilized in Adam, the new bias correction term Equation (47) can completely correct the effect of the initial value, making the initial value meaningless. Therefore, the first-order Taylor expansion of Equation (47) at $\beta_2^t = 0$ is considered:

$$\hat{v}_t = v_t + \beta_2^t(v_t - v_0^*) \tag{48}$$

On the one hand, when a sufficient number of samples are obtained, that is, $t \to +\infty$ and $\beta_2^t \to 0^+$, Equation (48) approaches Equation (47), which is little affected by the initial value. On the other hand, during the initial timesteps, the estimates of

Equation (48) will be biased towards the prior initial value vector, making the estimates more robust.

For the training process, the robustness of the adaptive estimates during the initial timesteps is more important due to the uncertainty of the sampler initialization. Thus, in addition to the variances, considering the initial value vector as $m_0^* = 0$, the bias correction term $\hat{m}_t \leftarrow m_t/(1 - \beta_1^t)$ of the 1st moment vector in Algorithm 2 can also be similarly changed into:

$$\hat{m}_t = m_t * (1 + \beta_1^t) \tag{49}$$

By introducing the prior initial value vector $v_0^*$ and taking Equations (48)-(49) as the initialization bias correction term, the adaptive estimate process is further updated as Algorithm 3 and it is utilized in the adaptive estimation of the variances $\sigma_i^2, i = 1, \cdots, D$ of each component of the parameter samples.

---

**Algorithm 3:** Second moment estimate with prior initial value vector. All operations on vectors are elementwise.

---

**Require:** $\beta_1, \beta_2 \in [0,1)$: Exponential decay rates for the moment estimates
**Require:** $K$: Batch size of samples obtained at each timestep
**Require:** $v_0^*$: Prior initial value vector of the 2nd moment vector
**Require:** $\{y_{1,k}\}_{k=1}^K, \cdots, \{y_{T,k}\}_{k=1}^K$: Samples obtained at subsequent timesteps $t = 1, \cdots, T$

  $m_0 \leftarrow 0$ (Initialize 1st moment vector)
  $v_0 \leftarrow v_0^*$ (Initialize 2nd moment vector)
  $t \leftarrow 0$ (Initialize timestep)
  **for** $t' = 1, \cdots, T$ **do**
    $t \leftarrow t + 1$
    $\bar{y}_t \leftarrow \frac{1}{K}\sum_{k=1}^K y_{t,k}$
    $m_t \leftarrow \beta_1 m_{t-1} + (1 - \beta_1)\bar{y}_t$ (Update biased 1st moment estimate)
    **if** *training process* **do**
      $\hat{m}_t \leftarrow m_t * (1 + \beta_1^t)$ (Compute robust bias-corrected 1st moment estimate)
    **else if** *testing process* **do**
      $\hat{m}_t \leftarrow m_t/(1 - \beta_1^t)$ (Compute bias-corrected 1st moment estimate)
    **end if**
    **if** $t = 1$ **do**
      $\hat{m}_0 \leftarrow \hat{m}_1$

**end if**
$v'_{t-1} \leftarrow (\hat{m}_t - \hat{m}_{t-1})^2 + v_{t-1}$
$\overline{y_t^2} \leftarrow \frac{1}{K}\left[(\hat{m}_t - \hat{m}_{t-1})^2 + \sum_{k=1}^{K}(y_{t,k} - \hat{m}_t)^2\right]$
$v_t \leftarrow \beta_2 v'_{t-1} + (1 - \beta_2)\overline{y_t^2}$ (Update biased centered 2nd moment estimate)
$\hat{v}_t \leftarrow v_t + \beta_2^t(v_t - v_0^*)$ (Compute robust bias-corrected centered 2nd moment estimate)
**end for**
**return** $\hat{v}_t$ at timesteps $t = 1, \cdots, T$ (Estimating the first two moments)

To more intuitively show the difference between Equation (47) and Equation (48) during the initial timesteps, the estimate for the first timestep $t = 1$ is calculated below. According to Algorithm 3, when $t = 1$, we have $v'_0 = v_0 = v_0^*$ determined by the prior initial value vector $v_0^*$ and $\overline{y_1^2} = \frac{1}{K}\sum_{k=1}^{K}\left(y_{1,k} - \frac{1}{K}\sum_{k'=1}^{K} y_{1,k'}\right)^2$ calculated based on the obtained samples $\{y_{1,k}\}_{k=1}^{K}$, then we have $v_1 = \beta_2 v'_0 + (1 - \beta_2)\overline{y_1^2} = \beta_2 v_0^* + (1 - \beta_2)\overline{y_1^2}$.

The bias-corrected centered second moment estimate computed by Equations (47) and (48) are shown in equations (50) and (51), respectively:

$$\hat{v}_1 = \overline{y_1^2} \tag{50}$$

$$\hat{v}_1 = \beta_2^2 v_0^* + (1 - \beta_2^2)\overline{y_1^2} \tag{51}$$

It can be seen that Equation (47) is not affected by the prior initial value vector $v_0^*$ at all, while Equation (48) gives a large weight $\beta_2^2$ to the prior $v_0^*$ at the first timestep.

# Appendix B *Simple proof of scale-invariance of the meta-learning procedure in AM-SGHMC*

The scale-invariance with respect to the posterior PDF mentioned in the main text means that when each component of the posterior PDF is linearly scaled by $\lambda_i, i = 1, \cdots, D$ and shifted by $b_i, i = 1, \cdots, D$ (the shape of the posterior PDF in each dimension is thus unchanged), the displacement between adjacent samples generated by the sampler is also correspondingly scaled, so that the sampling effect remains invariant.

Here the model parameters before and after scaling are represented as random variables $\boldsymbol{\Theta}^+$ and $\boldsymbol{\Theta}'$, respectively, with a relation $\theta'_i = \lambda_i \theta_i^+ + b_i, i = 1, \cdots, D$. For any $\boldsymbol{\theta}^+$ in the sample space of $\boldsymbol{\Theta}^+$, the corresponding $\boldsymbol{\theta}'$ is defined by $\theta'_i = \lambda_i \theta_i^+ + b_i, i = 1, \cdots, D$. According to the multivariate change of variables theorem in probability theory, which is detailed in Appendix C, the relationship between their PDFs $p_{\boldsymbol{\Theta}^+}(\boldsymbol{\theta}^+)$ and $p_{\boldsymbol{\Theta}'}(\boldsymbol{\theta}')$ is:

$$p_{\boldsymbol{\Theta}'}(\boldsymbol{\theta}') = \frac{1}{\prod_{i=1}^D \lambda_i} p_{\boldsymbol{\Theta}^+}(\boldsymbol{\theta}^+) \tag{52}$$

Considering that the potential energy $U(\boldsymbol{\theta}) = -\log p_{\boldsymbol{\Theta}}(\boldsymbol{\theta}) + c$ and $\frac{1}{\prod_{i=1}^D \lambda_i}$ is a constant, the relationship between their potential energy $U^+(\boldsymbol{\theta}^+)$ and $U'(\boldsymbol{\theta}')$ is:

$$\begin{aligned} U'(\boldsymbol{\theta}') &= -\log p_{\boldsymbol{\Theta}'}(\boldsymbol{\theta}') + c \\ &= -\log p_{\boldsymbol{\Theta}^+}(\boldsymbol{\theta}^+) - \log \frac{1}{\prod_{i=1}^D \lambda_i} + c \\ &= -\log p_{\boldsymbol{\Theta}^+}(\boldsymbol{\theta}^+) + c' \\ &= U^+(\boldsymbol{\theta}^+) + c'' \end{aligned} \tag{53}$$

i.e., $U'(\boldsymbol{\theta}') = U^+(\boldsymbol{\theta}^+) + c''$. Thus, $\mu_{U'} = \mu_{U^+} + c''$, $\sigma_{U'} = \sigma_{U^+}$ and subsequently, the normalized potential energy inputs have the relationship:

$$\begin{aligned} \widehat{U}'(\boldsymbol{\theta}') &= \frac{U'(\boldsymbol{\theta}') - \mu_{U'}}{\sqrt{2D}\sigma_{U'}} \\ &= \frac{(U^+(\boldsymbol{\theta}^+) + c'') - (\mu_{U^+} + c'')}{\sqrt{2D}\sigma_{U^+}} \\ &= \frac{U^+(\boldsymbol{\theta}^+) - \mu_{U^*}}{\sqrt{2D}\sigma_{U^+}} \\ &= \widehat{U}^+(\boldsymbol{\theta}^+) \end{aligned} \tag{54}$$

i.e., $\widehat{U}'(\boldsymbol{\theta}') = \widehat{U}^+(\boldsymbol{\theta}^+)$. Thus, $\partial_{\theta_i'}\widehat{U}'(\boldsymbol{\theta}') = \partial_{\theta_i'}\widehat{U}^+(\boldsymbol{\theta}^+) = \frac{1}{\lambda_i}\partial_{\theta_i^+}\widehat{U}^+(\boldsymbol{\theta}^+)$.

Considering that the relationship between standard deviations of $\Theta_i'$ and $\Theta_i^+$ is $\sigma_i' = \lambda_i \sigma_i^+$ due to the scaling, the normalized potential energy gradient inputs have the relationship:

$$\partial_{\theta_i'}\widehat{U}^{*'}(\boldsymbol{\theta}') = \sigma_i' * \partial_{\theta_i'}\widehat{U}'(\boldsymbol{\theta}')$$

$$= \lambda_i \sigma_i^+ * \frac{1}{\lambda_i}\partial_{\theta_i^+}\widehat{U}^+(\boldsymbol{\theta}^+)$$

$$= \sigma_i^+ * \partial_{\theta_i^+}\widehat{U}^+(\boldsymbol{\theta}^+)$$

$$= \partial_{\theta_i^+}\widehat{U}^{*+}(\boldsymbol{\theta}^+) \tag{55}$$

i.e., $\partial_{\theta_i'}\widehat{U}^{*'}(\boldsymbol{\theta}') = \partial_{\theta_i^+}\widehat{U}^{*+}(\boldsymbol{\theta}^+)$.

Considering that the auxiliary variables $\boldsymbol{p} \sim \mathcal{N}(\boldsymbol{0}, \boldsymbol{I})$ is invariant during scaling, i.e., the corresponding auxiliary variables before and after scaling will keep $\boldsymbol{p}' = \boldsymbol{p}^+$, it has been shown that the inputs of both networks $f_{\phi_Q}$ and $f_{\phi_D}$ remain unchanged after the scaling when the corresponding $\boldsymbol{z}' = (\boldsymbol{\theta}', \boldsymbol{p}')$ is used.

Then the outputs can be derived based on Equations (16)-(17):

$$\boldsymbol{f}'_{\phi_Q,i}(\boldsymbol{z}') = \sigma_i' * \left(c_1 + f_{\phi_Q}\big(\widehat{U}'(\boldsymbol{\theta}'), \boldsymbol{p}_i', Cate_i\big)\right)$$

$$= \lambda_i \sigma_i^+ * \left(c_1 + f_{\phi_Q}\big(\widehat{U}^+(\boldsymbol{\theta}^+), \boldsymbol{p}_i^+, Cate_i\big)\right)$$

$$= \lambda_i \boldsymbol{f}^+_{\phi_Q,i}(\boldsymbol{z}^+) \tag{56}$$

$$\boldsymbol{f}'_{\phi_D,i}(\boldsymbol{z}') = c_2 + f_{\phi_D}\left(\widehat{U}'(\boldsymbol{\theta}'), \boldsymbol{p}_i', \partial_{\theta_i'}\widehat{U}^{*'}(\boldsymbol{\theta}'), Cate_i\right)$$

$$= c_2 + f_{\phi_D}\left(\widehat{U}^+(\boldsymbol{\theta}^+), \boldsymbol{p}_i^+, \partial_{\theta_i^+}\widehat{U}^{*+}(\boldsymbol{\theta}^+), Cate_i\right)$$

$$= \boldsymbol{f}^+_{\phi_D,i}(\boldsymbol{z}^+) \tag{57}$$

i.e., $\boldsymbol{f}'_{\phi_Q,i}(\boldsymbol{z}') = \lambda_i \boldsymbol{f}^+_{\phi_Q,i}(\boldsymbol{z}^+)$ and $\boldsymbol{f}'_{\phi_D,i}(\boldsymbol{z}') = \boldsymbol{f}^+_{\phi_D,i}(\boldsymbol{z}^+)$, where $\boldsymbol{z}^+ = (\boldsymbol{\theta}^+, \boldsymbol{p}^+)$.

Now, we focus on one sampling step before scaling in each dimension $i$ at the discretized time $t$. Denoting a sample generated from standard Gaussian distribution $\mathcal{N}(0,1)$ as $\epsilon_0$, the $i$-th component of $\boldsymbol{\epsilon}_t \sim \mathcal{N}(\mathbf{0}, 2\eta C(\mathbf{z}_t))$ can be written as $\epsilon_{t,i} = \sqrt{2\eta f_{\phi_D,i}(\mathbf{z}_t)}\epsilon_0$. Starting with state $\mathbf{z}_t^+ = (\boldsymbol{\theta}_t^+, \mathbf{p}_t^+)$, according to Equations (18)-(19), this sampling step can be written as:

$$p_{t+1,i}^+ = \left(1 - \eta f_{\phi_D,i}^+(\mathbf{z}_t^+)\right)p_{t,i}^+ - \eta f_{\phi_Q,i}^+(\mathbf{z}_t^+)\frac{\partial U^+(\boldsymbol{\theta}_t^+)}{\partial \theta_{t,i}^+}$$

$$+ \eta \left(\frac{\partial f_{\phi_Q,i}^+(\mathbf{z}_t^+)}{\partial \theta_{t,i}^+} + \frac{\partial f_{\phi_D,i}^+(\mathbf{z}_t^+)}{\partial p_{t,i}^+}\right) + \sqrt{2\eta f_{\phi_D,i}^+(\mathbf{z}_t^+)}\epsilon_0 \qquad (58)$$

$$\theta_{t+1,i}^+ = \theta_{t,i}^+ + \eta f_{\phi_Q,i}^+(\hat{\mathbf{z}}_t^+)p_{t+1,i}^+ - \eta \frac{\partial f_{\phi_Q,i}^+(\hat{\mathbf{z}}_t^+)}{\partial p_{t+1,i}^+} \qquad (59)$$

where $\hat{\mathbf{z}}_t^+ = (\boldsymbol{\theta}_t^+, \mathbf{p}_{t+1}^+)$, and the new state $\mathbf{z}_{t+1}^+ = (\boldsymbol{\theta}_{t+1}^+, \mathbf{p}_{t+1}^+)$ is generated.

As for the sampling step after scaling, starting with the corresponding state $\mathbf{z}_t' = (\boldsymbol{\theta}_t', \mathbf{p}_t')$, i.e., $\theta_{t,i}' = \lambda_i \theta_{t,i}^+ + b_i$ and $\mathbf{p}_t' = \mathbf{p}_t^+$, it can be correspondingly written as:

$$p_{t+1,i}' = \left(1 - \eta f_{\phi_D,i}'(\mathbf{z}_t')\right)p_{t,i}' - \eta f_{\phi_Q,i}'(\mathbf{z}_t')\frac{\partial U'(\boldsymbol{\theta}_t')}{\partial \theta_{t,i}'}$$

$$+ \eta \left(\frac{\partial f_{\phi_Q,i}'(\mathbf{z}_t')}{\partial \theta_{t,i}'} + \frac{\partial f_{\phi_D,i}'(\mathbf{z}_t')}{\partial p_{t,i}'}\right) + \sqrt{2\eta f_{\phi_D,i}'(\mathbf{z}_t')}\epsilon_0 \qquad (60)$$

$$\theta_{t+1,i}' = \theta_{t,i}' + \eta f_{\phi_Q,i}'(\hat{\mathbf{z}}_t')p_{t+1,i}' - \eta \frac{\partial f_{\phi_Q,i}'(\hat{\mathbf{z}}_t')}{\partial p_{t+1,i}'} \qquad (61)$$

where $\hat{\mathbf{z}}_t' = (\boldsymbol{\theta}_t', \mathbf{p}_{t+1}')$.

For Equation (60), it can be further written as:

$$p_{t+1,i}' = \left(1 - \eta f_{\phi_D,i}^+(\mathbf{z}_t^+)\right)p_{t,i}^+ - \eta \lambda_i f_{\phi_Q,i}^+(\mathbf{z}_t^+)\frac{1}{\lambda_i}\frac{\partial (U^+(\boldsymbol{\theta}_t^+) + c'')}{\partial \theta_{t,i}^+}$$

$$+ \eta \left(\frac{1}{\lambda_i}\frac{\partial \left(\lambda_i f_{\phi_Q,i}^+(\mathbf{z}_t^+)\right)}{\partial \theta_{t,i}^+} + \frac{\partial f_{\phi_D,i}^+(\mathbf{z}_t^+)}{\partial p_{t,i}^+}\right) + \sqrt{2\eta f_{\phi_D,i}^+(\mathbf{z}_t^+)}\epsilon_0$$

$$= \left(1 - \eta f^+_{\phi_D,i}(z_t^+)\right) p^+_{t,i} - \eta f^+_{\phi_Q,i}(z_t^+) \frac{\partial(U^+(\boldsymbol{\theta}_t^+))}{\partial \theta^+_{t,i}}$$

$$+ \eta \left( \frac{\partial f^+_{\phi_Q,i}(z_t^+)}{\partial \theta^+_{t,i}} + \frac{\partial f^+_{\phi_D,i}(z_t^+)}{\partial p^+_{t,i}} \right) + \sqrt{2\eta f^+_{\phi_D,i}(z_t^+)} \epsilon_0$$

$$= p^+_{t+1,i} \tag{62}$$

i.e., $p'_{t+1,i} = p^+_{t+1,i}$ for $i = 1, \cdots, D$, that is, $\boldsymbol{p}'_{t+1} = \boldsymbol{p}^+_{t+1}$. That means $\hat{\boldsymbol{z}}'_t = (\boldsymbol{\theta}'_t, \boldsymbol{p}'_{t+1})$ is also the corresponding state of $\hat{\boldsymbol{z}}^+_t = (\boldsymbol{\theta}^+_t, \boldsymbol{p}^+_{t+1})$.

Thus, for Equation (61), it can be further written as:

$$\theta'_{t+1,i} - \theta'_{t,i} = \eta \lambda_i f^+_{\phi_Q,i}(\hat{z}_t^+) p^+_{t+1,i} - \eta \frac{\partial \left( \lambda_i f^+_{\phi_Q,i}(\hat{z}_t^+) \right)}{\partial p^+_{t+1,i}}$$

$$= \lambda_i \left( \eta f^+_{\phi_Q,i}(\hat{z}_t^+) p^+_{t+1,i} - \eta \frac{\partial f^+_{\phi_Q,i}(\hat{z}_t^+)}{\partial p^+_{t+1,i}} \right)$$

$$= \lambda_i \left( \theta^+_{t+1,i} - \theta^+_{t,i} \right) \tag{63}$$

i.e., $\theta'_{t+1,i} - \theta'_{t,i} = \lambda_i(\theta^+_{t+1,i} - \theta^+_{t,i})$. By substituting $\theta'_{t,i} = \lambda_i \theta^+_{t,i} + b_i$ into it, we can get $\theta'_{t+1,i} = \lambda_i \theta^+_{t+1,i} + b_i$. That means $\boldsymbol{z}'_{t+1} = (\boldsymbol{\theta}'_{t+1}, \boldsymbol{p}'_{t+1})$ is also the corresponding state of $\boldsymbol{z}^+_{t+1} = (\boldsymbol{\theta}^+_{t+1}, \boldsymbol{p}^+_{t+1})$.

According to the derivations above, when the posterior PDF is linearly scaled, starting from the initial state $\boldsymbol{z}'_0 = (\boldsymbol{\theta}'_0, \boldsymbol{p}'_0)$ corresponding to $\boldsymbol{z}^+_0 = (\boldsymbol{\theta}^+_0, \boldsymbol{p}^+_0)$ and using a fixed random seed, parameter samples $\boldsymbol{\theta}'_t$ corresponding to $\boldsymbol{\theta}^+_t$ can be generated by performing Equations (60)-(61) iteratively from a Markov chain.

Obviously, for any $\boldsymbol{\theta}'_0$ in the sample space of $\boldsymbol{\Theta}'$, there is always a $\boldsymbol{\theta}^+_0$ in the sample space of $\boldsymbol{\Theta}^+$ calculated by $\theta^+_{0,i} = \frac{\theta'_{0,i} - b_i}{\lambda_i}, i = 1, \cdots, D$, so that $\boldsymbol{\theta}'_0$ corresponds to $\boldsymbol{\theta}^+_0$. Since the initial state of AM-SGHMC can be arbitrary, it is proved that the input/output processing proposed in Section 3.1 has the scale-invariance with respect to the posterior PDF.

Considering that the shape of the conditional posterior PDF for parameters of the same category is relatively similar, it is feasible for parameters of same category $Cate_i$ to share one pair of meta-strategy networks $f_{\phi_Q}(\cdot,\cdot,Cate_i)$ and $f_{\phi_D}(\cdot,\cdot,\cdot,Cate_i)$.

**Appendix C** *Derivation of the potential energy after transformation*

According to the multivariate change of variables theorem in probability theory, suppose that $\boldsymbol{W}$ is a random variable taking values in $\mathcal{S} \subseteq \mathbb{R}^D$, and that $\boldsymbol{W}$ has a continuous distribution with PDF $p_W$. Suppose another random variable $\boldsymbol{\Theta} = r(\boldsymbol{W})$ where $r$ is a one-to-one and differentiable function from $\mathcal{S}$ onto $\mathcal{T} \subseteq \mathbb{R}^D$. Then the PDF $p_\Theta$ of $\boldsymbol{\Theta}$ is given by:

$$p_\Theta(\boldsymbol{\theta}) = p_W(\boldsymbol{w}) \left| \det\left(\frac{d\boldsymbol{w}}{d\boldsymbol{\theta}}\right) \right| \tag{64}$$

where $\left(\frac{d\boldsymbol{w}}{d\boldsymbol{\theta}}\right)$ is the first derivative of the inverse function $\boldsymbol{w} = r^{-1}(\boldsymbol{\theta})$ defined as the $D \times D$ matrix of first partial derivatives:

$$\left(\frac{d\boldsymbol{w}}{d\boldsymbol{\theta}}\right)_{i,j} = \frac{\partial w_i}{\partial \theta_j} \tag{65}$$

and $\det\left(\frac{d\boldsymbol{w}}{d\boldsymbol{\theta}}\right)$ is the determinant of the first derivative matrix called Jacobian of the inverse function $r^{-1}$.

Back to the transformation between states $\boldsymbol{\theta}$ and parameters $\boldsymbol{w}$ in the main text, where $p_W$ is known as the posterior PDF $p_W(\boldsymbol{w}) = p(\boldsymbol{w}|\mathcal{D})$, and the inverse function $r^{-1}$ is defined by Equations (27)-(28):

$$w_i = r_i^{-1}(\theta_i) = \begin{cases} f(\theta_i; b_{i,1}, \delta_{i,1}), & \theta_i < b_{i,1} \\ \theta_i, & b_{i,1} \leq \theta_i \leq b_{i,2} \\ f(\theta_i; b_{i,2}, \delta_{i,2}), & \theta_i > b_{i,2} \end{cases} \tag{66}$$

where the function $f(\theta; b, \delta)$ of $\theta$ is defined as:

$$f(\theta; b, \delta) = 2\delta \cdot \text{Sigmoid}\left(\frac{2(\theta - b)}{\delta}\right) + b - \delta \tag{67}$$

Since the inverse function $r^{-1}$ is elementwise and monotonically increasing in each component, the first derivative matrix $\left(\frac{d\boldsymbol{w}}{d\boldsymbol{\theta}}\right)$ becomes a positive definite diagonal matrix and the absolute value of Jacobian in Equation (64) can be simplified to:

$$\left|\det\left(\frac{d\boldsymbol{w}}{d\boldsymbol{\theta}}\right)\right| = \prod_{i=1}^{D}\frac{dw_i}{d\theta_i} \tag{68}$$

and thus Equation (64) becomes:

$$p_{\boldsymbol{\theta}}(\boldsymbol{\theta}) = p_W(\boldsymbol{w})\prod_{i=1}^{D}\frac{dw_i}{d\theta_i} \tag{69}$$

Considering the relationship between potential energy $U(\boldsymbol{\theta})$ and target PDF $p_{\boldsymbol{\theta}}(\boldsymbol{\theta})$: $U(\boldsymbol{\theta}) = -\log p_{\boldsymbol{\theta}}(\boldsymbol{\theta}) + c$, we take the negative logarithm of both sides of Equation (69) and get:

$$U(\boldsymbol{\theta}) = U(\boldsymbol{w}) - \sum_{i=1}^{D}\log\left(\frac{dw_i}{d\theta_i}\right) + c' \tag{70}$$

where $U(\boldsymbol{w})$ can be calculated following Equation (30). Denoting $\log\left(\frac{dw_i}{d\theta_i}\right)$ as $T(\theta_i)$, Equation (29) can be obtained.

Then we focus on the calculation of $T(\theta_i)$. Recalling Equation (66), we get:

$$T(\theta_i) = \log\left(\frac{dw_i}{d\theta_i}\right) = \log\left(\frac{dr_i^{-1}(\theta_i)}{d\theta_i}\right)$$

$$= \begin{cases} \log[f'(\theta_i; b_{i,1}, \delta_{i,1})], & \theta_i < b_{i,1} \\ 0, & b_{i,1} \leq \theta_i \leq b_{i,2} \\ \log[f'(\theta_i; b_{i,2}, \delta_{i,2})], & \theta_i > b_{i,2} \end{cases} \tag{71}$$

Denoting $\log[f'(\theta; b, \delta)]$ as $g(\theta; b, \delta)$, Equation (31) can be obtained.

Finally, after calculating the first derivative $f'(\theta; b, \delta) = \frac{df(\theta; b, \delta)}{d\theta}$ of function $f(\theta; b, \delta)$ defined as Equation (67), Equation (32) can be obtained:

$$g(\theta; b, \delta) = \log\left[\frac{df(\theta; b, \delta)}{d\theta}\right]$$

$$= 2\log Sigmoid\left(\frac{2(\theta - b)}{\delta}\right) - \frac{2(\theta - b)}{\delta} + \log 4 \qquad (72)$$

Now the potential energy after transformation is fully derived.

## Appendix D *Network architecture and training setup*

Thanks to the scale-invariant input/output processing in Section 3.1, the meta-strategy for the sampler for various Bayesian updating problems with the same type of structure can be learned well by simple neural networks embedded in AM-SGHMC. The architecture of the meta-strategy network is shown in Figure 16, which can be divided into three parts according to the dashed boxes.

Parts I and III are additional input/output processing to make it suitable for network training. In part I, inputs $\widehat{U}(\boldsymbol{\theta})$, $p_i$, $\partial_{\theta_i}\widehat{U}^*(\boldsymbol{\theta})$ and $Cate_i$ are converted into network inputs $\boldsymbol{i}_U$, $\boldsymbol{i}_{p_i}$, $\boldsymbol{i}_{G_i}$ and $\boldsymbol{i}_{C_i}$ by functions:

$$\boldsymbol{i}_U = \log\left[\left(ReLU(\widehat{U}(\boldsymbol{\theta}) + 1)\right)^2 + e - 1\right] - 1 \qquad (73)$$

$$\boldsymbol{i}_{p_i} = 3\, Sigmoid\left(\frac{p_i}{10}\right) - 1.5 \qquad (74)$$

$$\boldsymbol{i}_{G_i} = 3\, Sigmoid\left(\frac{\partial_{\theta_i}\widehat{U}^*(\boldsymbol{\theta})}{30}\right) - 1.5 \qquad (75)$$

$$\boldsymbol{i}_{C_i} = one\_hot(Cate_i) \qquad (76)$$

where $ReLU(\cdot)$ and $Sigmoid(\cdot)$ are two activation functions widely used in neural networks, and $one\_hot(\cdot)$ is one-hot encoding, one of the most important encoding techniques for categorical data. In the shear-building example, the model parameters are divided into three categories: stiffness for $k_i$, damping for $c_i$, and noise level for $\sigma$. In the braced-frame building example, they are also divided into three categories:

outer modulus for $E_i^{(j)}, j = 1,2,3,4$, inner modulus for $E_i^{(5)}$, and noise level for $\sigma$. In Part III, the network outputs $\boldsymbol{o}_Q$ and $\boldsymbol{o}_D$ are converted into function outputs by:

$$f_{\phi_Q}(\hat{U}(\boldsymbol{\theta}), p_i, Cate_i) = M_Q\, Sigmoid(5\boldsymbol{o}_Q) \tag{77}$$

$$f_{\phi_D}(\hat{U}(\boldsymbol{\theta}), p_i, \partial_{\theta_i}\hat{U}^*(\boldsymbol{\theta}), Cate_i) = M_D\, Sigmoid(5\boldsymbol{o}_D) \tag{78}$$

where the maximum values of functions $f_{\phi_Q}$ and $f_{\phi_D}$ are controlled by the parameters $M_Q$ and $M_D$. After setting step-size $\eta$, we recommend setting the parameters $M_Q$ and $M_D$ to satisfy $\eta M_Q \approx 3$ and $\eta M_D \lesssim 1$, respectively. For example, we set $\eta = \sqrt{0.001}$, $M_Q = 100$ and $M_D = 30$.

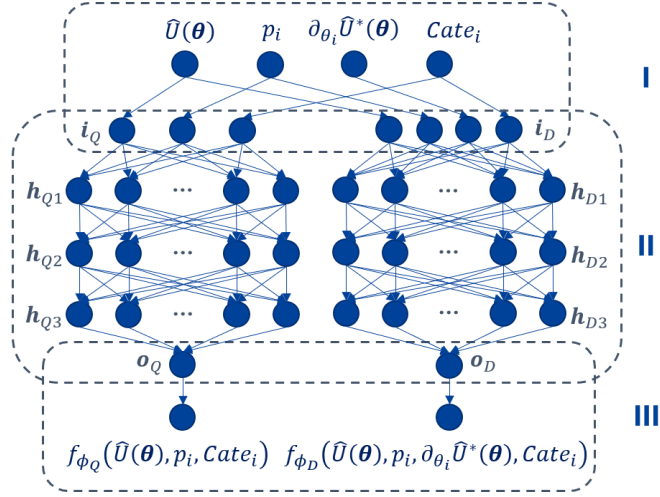

**Figure 16.** Schematic of meta-strategy network architecture.

Part II is two neural networks embedded in functions $f_{\phi_Q}$ and $f_{\phi_D}$, respectively. In the shear-building example, both of them are 3-hidden-layer multilayer perceptrons (MLPs) with 10 units for each hidden layer, and Leaky ReLU is utilized as the activation function for each unit of the hidden layers. In the braced-frame building example, due to the complexity of the posterior PDF, the network input-output ($\boldsymbol{i}_Q$-$\boldsymbol{o}_Q$ and $\boldsymbol{i}_D$-$\boldsymbol{o}_D$) relationship will change rapidly during the adaptive estimates, so that the slow speed of the network training will slow down, or even disrupt, the process of adaptive estimates. In order to make the networks more flexible in the rapidly changing environment, we add a shortcut connection beside the 3-hidden-layer MLP (NN), as shown in Figure 17. In the added shortcut connection, the linear transform part (Lin) and the radial basis function part (RBFs) can respond quickly to global and local changes in the $\boldsymbol{i}$-$\boldsymbol{o}$ relationship, respectively. They will be trained only during the adaptive estimate process and then fixed to remain unchanged. Note that the added short-cut part in Example II is also applicable for Example I.

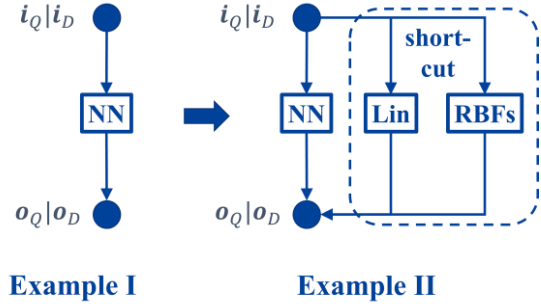

**Figure 17.** Schematic of Part II networks for the two examples.

For the training process, $K_0 = 64$ parallel chains are simulated for 100 epochs and each one consists of 10 sub-epochs. After each epoch, states $\boldsymbol{z} = (\boldsymbol{\theta}, \boldsymbol{p})$ of the Markov chains are re-initialized using replay techniques with probability 0.2, that is, 20% of the states are re-initialized using the states simulated earlier. For each sub-epoch, the sampler is simulated 90 steps and then is updated by the Adam optimizer with learning rate of 0.01 and exponential decay rates $(\beta_1, \beta_2)_{\text{Adam}} = (0.5, 0.75)$. For each $T_T = 15$ steps, $K = 10$ chains are randomly selected to estimate the loss function $Loss_{t_0}\left(\{\{\boldsymbol{\theta}^k_{t_0+s}\}_{s=0}^{T_T}\}_{k=1}^{K}\right)$ as in Figure 3, and the gradient is backpropagated and accumulated. The adaptive estimates are only updated during the last 6 sub-epochs of the first 50 epochs with $(\beta_1, \beta_2)_{\boldsymbol{\theta}} = (0.99, 0.999)$ and $(\beta_1, \beta_2)_U = (0.99, 0.998)$ for parameter samples $\boldsymbol{\theta}$ and potential energy $U(\boldsymbol{\theta})$.